\pdfoutput=1

\documentclass[11pt,twoside,a4paper,cmspaper,final,collab]{cms-tdr}
\def\svnVersion{469405P}\def\cmsCernNoTag{CERN-EP-2017-328}\def\cmsCernDate{\today}\def\cmsMessage{Published in the European Physical Journal C as \href{http://dx.doi.org/10.1140/epjc/s10052-018-6049-9}{\doi{10.1140/epjc/s10052-018-6049-9.}}}

\begin{document}\cmsNoteHeader{SMP-16-018}

\hyphenation{had-ron-i-za-tion}
\hyphenation{cal-or-i-me-ter}
\hyphenation{de-vices}
\RCS$Revision: 469388 $
\RCS$HeadURL: svn+ssh://svn.cern.ch/reps/tdr2/papers/SMP-16-018/trunk/SMP-16-018.tex $
\RCS$Id: SMP-16-018.tex 469388 2018-07-23 08:26:17Z paoloa $
\newlength\cmsFigWidth
\ifthenelse{\boolean{cms@external}}{\setlength\cmsFigWidth{0.85\columnwidth}}{\setlength\cmsFigWidth{0.45\textwidth}}
\ifthenelse{\boolean{cms@external}}{\providecommand{\cmsLeft}{top\xspace}}{\providecommand{\cmsLeft}{left\xspace}}
\ifthenelse{\boolean{cms@external}}{\providecommand{\cmsRight}{bottom\xspace}}{\providecommand{\cmsRight}{right\xspace}}

\providecommand{\ll}{\ensuremath{\ell\ell}\xspace}
\providecommand{\jj}{\ensuremath{\mathrm{jj}}\xspace}
\providecommand{\lljj}{\ensuremath{\ell\ell\mathrm{jj}}\xspace}
\providecommand{\Zjj}{\ensuremath{\PZ\mathrm{jj}}\xspace}
\providecommand{\ewklljj}{\ensuremath{\mathrm{EW}\,\ell\ell\mathrm{jj}}\xspace}
\providecommand{\ewkzjj}{\ensuremath{\mathrm{EW}\,\PZ\mathrm{jj}}\xspace}
\providecommand{\ewkgjj}{\ensuremath{\mathrm{EW}\,\gamma\mathrm{jj}}\xspace}
\providecommand{\gjj}{\ensuremath{\gamma\mathrm{jj}}\xspace}
\providecommand{\dyzjj}{\ensuremath{\mathrm{DY}\,\PZ\mathrm{jj}}\xspace}
\providecommand{\ptj}{\ensuremath{p_{\mathrm{T j}}}\xspace}
\providecommand{\ptjj}{\ensuremath{p_{\mathrm{T jj}}}\xspace}
\newcommand{\x}{\ensuremath{\phantom{0}}}
\newcommand{\y}{\ensuremath{\phantom{.}}}
\providecommand{\NA}{{\ensuremath{\text{---}}}}

\cmsNoteHeader{SMP-16-018}
\title{Electroweak production of two jets in association with a Z boson in proton-proton collisions at $\sqrt{s}= $ 13\TeV}
\titlerunning{Electroweak production of two jets in association with a Z boson in pp collisions at $\sqrt{s}=13\TeV$}

\date{\today}

\abstract{
A measurement of the electroweak (EW) production of two jets in association with a $\PZ$ boson
in proton-proton collisions at $\sqrt{s}=13\TeV$ is presented, based on data recorded in 2016 
by the CMS experiment at the LHC corresponding to an integrated luminosity of 35.9\fbinv.
The measurement is performed in the $\ell\ell\mathrm{jj}$ final state with $\ell$ including electrons
and muons, and the jets j corresponding to the quarks produced in the hard interaction.
The measured cross section in a kinematic region defined by invariant masses
$m_{\ell\ell} >50\GeV$, $m_{\mathrm{jj}} >120\GeV$,
and transverse momenta $p_{\mathrm{T j}} > 25\GeV$ is 
$\sigma_\mathrm{EW}(\ell\ell\mathrm{jj})= 534 \pm 20\stat \pm 57\syst\unit{fb}$,
in agreement with leading-order standard model predictions.
The final state is also used to perform a search for anomalous trilinear gauge couplings.
No evidence is found and limits on anomalous trilinear gauge couplings associated with dimension-six
operators are given in the framework of an effective field theory.
The corresponding 95\% confidence level intervals are 
$-2.6 <  c_{WWW}/\Lambda^2  < 2.6\TeV^{-2}$ and
$-8.4 <  c_{W}/\Lambda^2  < 10.1\TeV^{-2}$.
The additional jet activity of events in a signal-enriched region is also studied,
and the measurements are in agreement with predictions.
}

\hypersetup{%
pdfauthor={CMS Collaboration},%
pdftitle={Measurement of the electroweak production of two jets in association with a Z boson in proton-proton collisions at sqrt(s)=13 TeV},%
pdfsubject={CMS},%
pdfkeywords={CMS, physics, Z, vector boson fusion, electroweak interaction, cross section}}

\maketitle

\section{Introduction\label{sec:intro}}

In proton-proton (pp) collisions at the CERN LHC,
the production of dileptons ($\ell\ell$) consistent with the Z boson invariant mass 
in association with two jets (\jj)
is dominated by events where the dilepton pair is produced by
a Drell--Yan (DY) process, in association with jets from  strong interactions.
This production is governed by a mixture of electroweak (EW)
and strong processes of order
$\alpha_\mathrm{EW}^2\alpha_\mathrm{S}^2$, where
$\alpha_\mathrm{S}$ is the strong coupling
and $\alpha_\mathrm{EW}$ is the EW coupling strength.

The pure electroweak production of the \lljj final state,
at order $\alpha_\mathrm{EW}^4$, is less frequent~\cite{Oleari:2003tc},
and includes production via the vector boson fusion
(VBF) process, with its distinctive signature  of
two jets with both large energy and separation in pseudorapidity $\eta$.
In this paper the electroweak production is referred
to as EW \Zjj, and the 
two jets produced through the fragmentation of the outgoing quarks
are referred to as ``tagging jets''.

\begin{figure*}[htb] 
\centering
\includegraphics[width=0.25\textwidth]{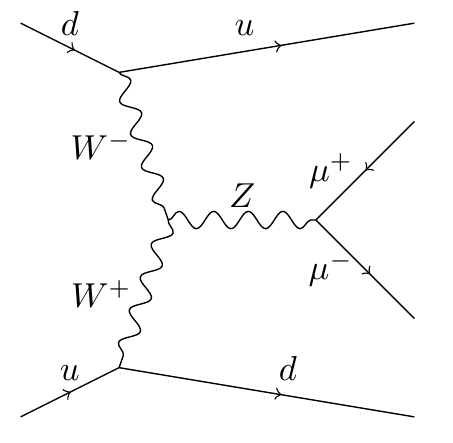} \hfil
\includegraphics[width=0.25\textwidth]{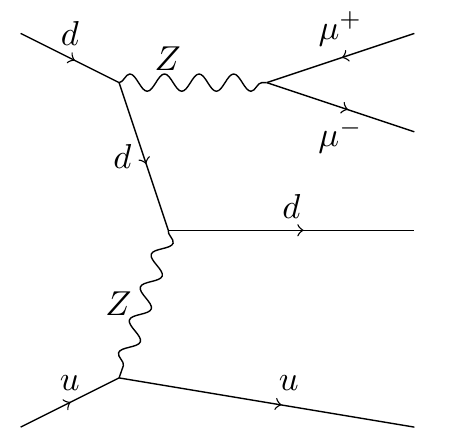} \hfil
\includegraphics[width=0.25\textwidth]{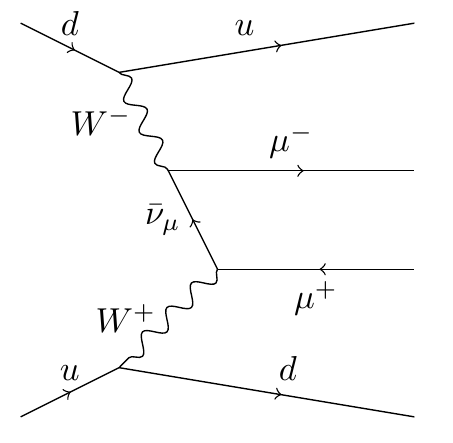}
\caption{
Representative Feynman diagrams for purely electroweak amplitudes for
dilepton production in association with two jets:
vector boson fusion (left),
bremsstrahlung-like (center),
and  multiperipheral production (right).
\label{fig:sigdiagram}}
\end{figure*}

Figure~\ref{fig:sigdiagram} shows representative Feynman diagrams for the EW \Zjj\ 
signal, namely VBF (left),
bremsstrahlung-like (center), and
multiperipheral (right) production.
Gauge cancellations lead to a large negative interference between the VBF 
process and the other two processes,
with the interferences from the bremsstrahlung-like production being larger.
Interference with multiperipheral 
production is limited to cases where the dilepton mass is close to
the Z boson peak mass~\cite{Chehime:1992ub}.

In the inclusive production of \lljj\ final states,
some of the nonexclusive EW interactions with identical
initial and final states can interfere with the exclusive EW interactions that are
shown in Fig.~\ref{fig:sigdiagram}.
This interference effect between the signal production and the main background processes
is much smaller than the interference effects among the EW production amplitudes,
but needs to be taken into account when measuring the signal 
contribution~\cite{Chatrchyan:2013jya,Khachatryan:2014dea}.

\begin{figure*}[htb] 
\centering
\includegraphics[width=0.25\textwidth]{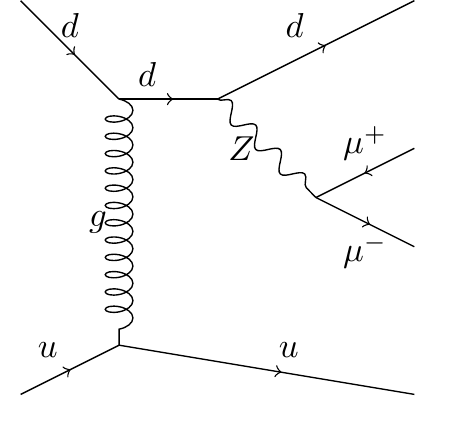} \hfil
\includegraphics[width=0.25\textwidth]{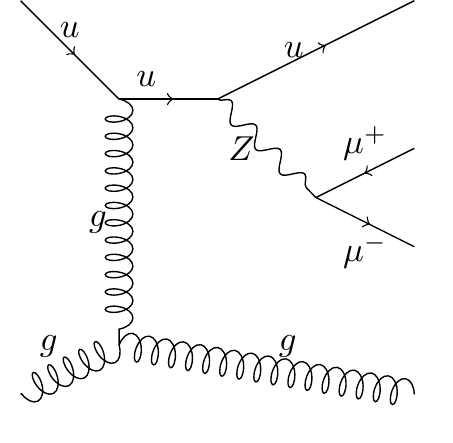}
\caption{Representative Feynman diagrams for order $\alpha_\mathrm{S}^2$
corrections to DY production that constitute the main background
for the measurement.}
\label{fig:bkgdiagram}
\end{figure*}

Figure~\ref{fig:bkgdiagram}~(left) shows
one example of corrections to order $\alpha_\mathrm{S}^2$ for DY production
that have the same initial and final states as those in
Fig.~\ref{fig:sigdiagram}.
A process at order $\alpha_\mathrm{S}^2$
that does not interfere with the EW signal is shown in
Fig.~\ref{fig:bkgdiagram}~(right).

The study of EW \Zjj\ processes is part of a more general
investigation of standard model (SM) vector boson fusion and scattering processes that include
studies of Higgs boson production~\cite{Aad:2012tfa,Chatrchyan:2012ufa,Chatrchyan:2013lba} and
searches for physics beyond the SM~\cite{Cho:2006sx,Dutta:2012xe,Khachatryan:2015kxa,Khachatryan:2016mbu}.
When isolated from the backgrounds, the properties
of EW \Zjj\ events can be compared with SM predictions.
Probing the additional hadronic activity in selected events
can shed light
on the modelling of additional parton radiation~\cite{Bjorken:1992er,Schissler:2013nga},
which is important for signal selection or vetoing of background events.

New physics could appear in the form of anomalous trilinear gauge couplings
(ATGCs)~\cite{Hagiwara:1993ck,Degrande:2012wf}
that can be parameterized with higher-dimensional operators.
Their measurements could provide an indirect search for
new physics at mass scales not directly accessible at the LHC.

At the LHC, the EW \Zjj\ process was first measured 
by the CMS experiment using pp collisions at
$\sqrt{s}=7\TeV$~\cite{Chatrchyan:2013jya},
and then at $\sqrt{s}=8\TeV$
by both the CMS~\cite{Khachatryan:2014dea} and ATLAS~\cite{Aad:2014dta} experiments.
The ATLAS experiment has also performed measurements at $\sqrt{s}=13\TeV$~\cite{Aaboud:2017emo},
with a data sample corresponding to an integrated luminosity of 3.2\fbinv. 
All results so far agree with the expectations of the SM
within a precision of approximately 20\%.

This paper presents a measurement with the CMS detector
using pp collisions collected at $\sqrt{s}=13\TeV$
during 2016, corresponding to an integrated luminosity of 35.9\fbinv.
A multivariate analysis,
based on the methods developed for the 7 and 8 TeV data  
results~\cite{Chatrchyan:2013jya,Khachatryan:2014dea},
is used to separate signal events from the large DY + jets background.
Analysis of the 13\TeV data with larger integrated luminosity and larger
predicted total cross section offers an opportunity to measure the cross section
at a higher energy and reduce the uncertainties of the earlier measurements.

Section~\ref{sec:cmsexperiment} describes the experimental
apparatus
and Section~\ref{sec:simulation} the event simulations.
Event selection procedures are described in Section~\ref{sec:evsel},
together with the
selection efficiencies and background models in control regions.
Section~\ref{sec:sigdisc} details the strategy adopted 
to extract the signal from the data, and the corresponding
systematic uncertainties are summarized in Section~\ref{sec:systunc}.
The cross section and anomalous coupling results are presented in Sections~\ref{sec:results}
and \ref{sec:atgc}, respectively.
Section~\ref{sec:hadactivity} provides
a study of the additional hadronic activity
in an EW \Zjj-enriched region.
Finally, a brief summary of the results is given in Section~\ref{sec:summary}.

\section{The CMS detector}
\label{sec:cmsexperiment}

The central feature of the CMS apparatus is a
superconducting solenoid of 6\unit{m} internal diameter,
providing a magnetic field of 3.8\unit{T}.
Within the solenoid volume are a silicon pixel and strip tracker,
a lead tungstate crystal electromagnetic calorimeter (ECAL),
and a brass and scintillator hadron calorimeter (HCAL),
each composed of a barrel and two endcap sections.
Forward calorimeters extend the $\eta$ coverage provided by the barrel and endcap detectors.
Muons are measured in gas-ionization detectors embedded in the steel flux-return yoke outside the solenoid.

The tracker measures charged particles within the pseudorapidity range $\abs{\eta} < 2.5$.
It consists of 1440 pixel and 15\,148 strip detector modules.
For nonisolated particles of $1 < \pt < 10\GeV$ and $\abs{\eta} < 1.4$,
the track resolutions are typically 1.5\% in \pt and 25--90 (45--150)\mum
in the transverse (longitudinal) impact parameter \cite{TRK-11-001}.

The electron momenta are estimated by combining energy measurements in the ECAL with
momentum measurements in the tracker~\cite{Khachatryan:2015hwa}.
The dielectron invariant mass resolution for $\PZ \rightarrow \Pe \Pe$ decays is 1.9\% when both electrons
are in the ECAL barrel, and 2.9\% when both electrons are in the endcaps.

Muons are measured in the pseudorapidity range $\abs{\eta} < 2.4$ with detection planes made using
three technologies:
drift tubes, cathode strip chambers, and resistive-plate chambers.
Matching muons to tracks measured in the silicon tracker results in a relative transverse momentum
resolution for muons with $20 <\pt < 100\GeV$ of 1.3--2.0\% in the barrel and better than 6\% in the endcaps.
The \pt resolution in the barrel is better than 10\% for muons with \pt up to 1\TeV~\cite{Chatrchyan:2012xi}.

The offline analysis uses reconstructed charged-particle tracks and candidates from
the particle-flow (PF) algorithm~\cite{Sirunyan:2017ulk}.
In the PF event reconstruction, all stable particles in the event, \ie electrons, muons, photons,
charged and neutral hadrons,
are reconstructed as PF candidates using information from all subdetectors to obtain an
optimal determination of their direction, energy, and type.
The PF candidates are then used to reconstruct the jets and the missing transverse momentum.

A more detailed description of the CMS detector, together with a
definition of the coordinate system and the relevant kinematic variables,
can be found in Ref.~\cite{Chatrchyan:2008zzk}.

\section{Simulation of signal and background events}
\label{sec:simulation}

{\tolerance=600
Signal events are simulated at leading order (LO) using
the \MGvATNLO~(v2.3.3) Monte Carlo (MC) generator~\cite{Alwall:2011uj,Alwall:2014hca},
interfaced with \PYTHIA (v8.212)~\cite{Sjostrand:2007gs,Sjostrand:2014zea}
for parton showering (PS) and hadronization.
The NNPDF30 (nlo\_as0130)~\cite{Ball:2014uwa} parton distribution functions (PDF) are
used to generate the events.
The underlying event (UE) is modelled 
using the CUETP8M1 tune~\cite{Khachatryan:2015pea}.
The simulation does not include extra partons at
matrix element (ME) level.
The signal is defined in the kinematic region with dilepton invariant mass $m_{\ell\ell} >50\GeV$,
parton transverse momentum $\ptj > 25\GeV$,
and diparton invariant mass $m_{\mathrm{jj}} >120\GeV$.
The cross section of the $\ell\ell\mathrm{jj}$ final state (with $\ell$ = e or $\mu$),
applying the above fiducial cuts, is calculated to be
$\sigma_\mathrm{LO}(\mathrm{EW}~\ell\ell\mathrm{jj})=543^{+7}_{-9}\,\text{(scale)}\pm22\,\text{(PDF)}\unit{fb}$,
where the first uncertainty is obtained by changing simultaneously
the factorization ($\mu_{\rm F}$) and renormalization ($\mu_{\rm R}$) scales
by factors of $2$ and $1/2$, and the second reflects the uncertainties
in the NNPDF30 PDFs.
The LO signal cross section and relevant kinematic distributions
estimated with \MGvATNLO are found to be in agreement within 5\%
with the next-to-leading order (NLO)
predictions of the \textsc{vbfnlo} generator (v.2.7.1)~\cite{Arnold:2008rz,Arnold:2011wj,Arnold:2012xn}
that includes NLO QCD corrections.
For additional comparisons, signal events have also been simulated with
the \HERWIGpp (v2.7.1)~\cite{Bahr:2008pv} PS, using the EE5C~\cite{Seymour:2013qka}  tune.
\par}

Events coming from processes including ATGCs are generated with the same setting
as the SM sample, but include additional information for reweighting
in a three-dimensional effective field theory (EFT) parameter space,
as described in more detail in Sec.~\ref{atgc:sec2}.

Background DY events are also simulated 
with \MGvATNLO using (i) an NLO ME calculation with up to three final-state
partons generated from quantum chromodynamics (QCD) interactions, and
(ii) an LO ME calculation with up to four partons. The ME-PS matching is
performed following the FxFx prescription~\cite{Frederix:2012ps} for the NLO case,
and the MLM prescription~\cite{Mangano:2006rw,Alwall:2007fs} for the LO case.
The NLO background simulation is used to extract the final results, while the
independent LO samples are used to perform the multivariate discriminant training.
The dilepton DY production for $m_{\ell\ell}>50\GeV$ is normalized
to $\sigma_\text{th}(\mathrm{DY})=5.765\unit{nb}$, which is computed
at next-to-next-to-leading order (NNLO)
with \textsc{fewz} (v3.1)~\cite{Melnikov:2006kv}.

The evaluation of the interference between \ewkzjj\ and \dyzjj\ processes
relies on predictions obtained with \MGvATNLO.
A dedicated sample of  events arising from the interference terms is generated directly
by selecting the contributions of order $\alpha_\mathrm{S}\alpha_\mathrm{EW}^3$, 
and passing them through the full detector simulation to estimate
the expected interference contribution.

Other backgrounds  are expected from other sources of events with
two opposite-sign and same-flavour leptons  together with jets.
Top quark pair events are
generated with \POWHEG (v2.0)~\cite{Nason:2004rx,Frixione:2007vw,Alioli:2010xd}
and normalized to the inclusive cross section calculated at NNLO together with
next-to-next-to-leading logarithmic
corrections~\cite{Kidonakis:2012db,Czakon:2013goa}.
Single top quark processes are modelled at NLO
with \POWHEG~\cite{Alioli:2010xd,Nason:2004rx,Frixione:2007vw,Alioli:2009je,Re:2010bp}
and normalized to cross sections of $71.7\pm2.0\unit{pb}$,
$217\pm3\unit{pb}$, and $10.32\pm0.20\unit{pb}$ respectively for
the tW, $t$-, and $s$-channel production~\cite{Kidonakis:2012db,Kidonakis:2013zqa}.
The diboson production processes $\PW\PW$,
$\PW\PZ$, and $\PZ\PZ$ are generated with \PYTHIA 
and normalized to NNLO cross section computations obtained with 
\MCFM (v8.0)~\cite{Campbell:2010ff}.
The abbreviation VV is used in this document when
referring to the sum of the processes that yield two vector bosons.

The contribution from diboson processes with
\lljj\ final states, such as ZW and ZZ,
to the signal definition is small,
and these contributions are not included in the background.

The production of a $\PW$ boson in association with jets,
where the $\PW$ decays to a charged lepton and a neutrino,
is also simulated with \MGvATNLO,
and normalized to a total cross section of 61.53\unit{nb}, computed at NNLO with \textsc{fewz}.
Multijet QCD processes are also studied in simulation, but are found to yield negligible contributions to the selected events.
All background productions make use of the \PYTHIA
PS model with the CUETP8M1 tune.

A detector simulation based on \GEANTfour (v.9.4p03)~\cite{Agostinelli:2002hh,Allison:2006ve}
is applied to all the generated signal and background samples.
The presence of multiple pp interactions in the same bunch crossing  (pileup)
is incorporated by simulating additional interactions
(both in-time and out-of-time with respect to the hard 
interaction) with a multiplicity
that matches the distribution observed in data.
The additional events are simulated with \PYTHIA (v8.212)
making use of the NNPDF23 (nlo\_as0130)~\cite{Ball:2012cx}
PDF, and the CUETP8M1 tune.
The average pileup is estimated to be about 27
additional interactions per bunch crossing.

\section{Reconstruction and selection of events}
\label{sec:evsel}
Events containing two isolated, high-\pt leptons, and at least two high-\pt jets are selected.
Isolated single-lepton triggers are used to acquire the data,
where the lepton is required to have $\pt>27\GeV$ for the electron trigger
and $\pt>24\GeV$ for the muon trigger~\cite{Khachatryan:2016bia}.

{\tolerance=1200
In the offline reconstruction, electrons are reconstructed from clusters of energy
deposits in the ECAL that match tracks extrapolated from the
silicon tracker~\cite{Khachatryan:2015hwa}.
Offline muons are reconstructed by fitting trajectories based on hits in the silicon
tracker and in the muon system~\cite{Chatrchyan:2012xi}.
Reconstructed electron or muon candidates
are required to have $\pt>20\GeV$.
Electron candidates are required to be reconstructed
within $\abs{\eta}\leq 2.4$, excluding barrel-to-endcap
$1.444 < |\eta| < 1.566$
transition regions of the ECAL~\cite{Chatrchyan:2008zzk}.
Muon candidates are required to be reconstructed in the fiducial
region $\abs{\eta}\leq 2.4$ of the muon system.
The track associated with a lepton candidate is required
to have both its transverse and longitudinal impact parameters
compatible with the position of the main primary vertex (PV) of the event.
The reconstructed PV with the largest value of summed physics-object $\pt^2$ is
taken to be the primary $\Pp\Pp$ interaction vertex.
The physics objects are the objects returned by a jet finding algorithm~\cite{Cacciari:2008gp,Cacciari:2011ma}
applied to all charged particle tracks associated with the vertex, plus the corresponding
associated missing transverse momentum.
\par}

The leptons are required to be isolated.
The isolation is calculated from particle candidates reconstructed by the PF algorithm
and is corrected for pileup on an event-by-event basis.
The sum of scalar \pt of all particle candidates
reconstructed in an isolation cone
with radius $R=\sqrt{\smash[b]{(\Delta \eta)^{2}+(\Delta \phi)^{2}}}=0.4$
around the momentum vector of the lepton is required to be below
15 (25)\% of the electron (muon) \pt value.
The two isolated leptons with opposite electric charge
and highest \pt are chosen to form the dilepton pair,
and are required to have $\pt>30\GeV$ and $\pt>20\GeV$ for the \pt-leading and subleading lepton,
respectively.
Events with additional leptons are kept in the event selection.
Same-flavour dileptons (ee or $\mu\mu$)
compatible with $\PZ\to\ell\ell$ decays are then selected by
requiring $|m_{\PZ}-m_{\ell\ell}|<15\GeV$,
where $m_{\PZ}$ is the mass  of the $\PZ$ boson~\cite{Olive:2016xmw}.

Jets are reconstructed by clustering PF candidates with the anti-\kt
algorithm~\cite{Cacciari:2005hq,Cacciari:2008gp} using a distance parameter of 0.4.
The jet momentum is determined as the vector sum of all particle momenta in the jet,
and is found from simulation to be within 5 to 10\% of the true momentum over the
whole \pt spectrum and detector acceptance~\cite{Sirunyan:2017ulk}.

{\tolerance=600
An offset correction is applied to jet
energies to take into account the contribution from additional proton-proton interactions
within the same or nearby bunch crossings. Jet energy corrections are derived from simulation,
and are confirmed with in situ measurements of the energy balance in dijet, multijet, 
photon + jet, and leptonically decaying $\PZ$ + jet events~\cite{Chatrchyan:2011ds}.
Loose jet identification criteria are applied to reject misconstructed jets resulting
from detector noise~\cite{CMS-PAS-JME-16-003}.
Loose criteria are also applied to remove jets heavily contaminated with pileup energy
(clustering of energy deposits not associated with a parton from the
primary $\Pp\Pp$ interaction)~\cite{CMS-PAS-JME-13-005,CMS-PAS-JME-16-003}.
The efficiency of the jet identification criteria is greater than 99\%, rejecting 90\%
of background pileup jets with $\pt\simeq 50\GeV$.
The jet energy resolution (JER) is typically ${\approx}15\%$ at 10\GeV,
8\% at 100\GeV, and 4\% at 1\TeV~\cite{Chatrchyan:2011ds}.
Jets reconstructed with $\pt\geq15\GeV$ and ${\abs{\eta}\leq4.7}$ are used in the analysis.
\par}

The two highest \pt jets are defined as the tagging jets,
and are required  to have $\pt>50\GeV$ and $\pt>30\GeV$
for the \pt-leading and subleading jet, respectively.
The invariant mass of the two tagging jets is required to satisfy $m_{\mathrm{jj}}>200\GeV$.

A multivariate analysis technique, described in Section~\ref{sec:sigdisc}, is used 
to provide an optimal separation of the \dyzjj\ and \ewkzjj\ components of the inclusive
\lljj\ spectrum.
The main discriminating variables are the 
dijet invariant mass $m_{\mathrm{jj}}$ and the pseudorapidity separation $\Delta\eta_{\mathrm{jj}}$.
Other variables used in the multivariate analysis are described below.

Table~\ref{tab:yields} reports the expected and observed
event yields after the initial  selection and after imposing a minimum
value for the final discriminator output that defines the signal-enriched
region used for the studies of additional hadronic activity described in Section~\ref{sec:hadactivity}.

\begin{table*}[htbp]
\centering
\topcaption{Event yields expected for background and signal
processes using the initial selections
and with a cut on the multivariate analysis output (BDT) that provides
signal $\approx$ background. The yields are
compared to the data observed in the different channels and categories.
The total uncertainties quoted for signal, \dyzjj,
dibosons, and processes with top
quarks (\ttbar and single top quarks)
include the simulation statistical uncertainty.\label{tab:yields}}
\renewcommand{\arraystretch}{1.05}
\begin{tabular}{l|cc|cc}
\multirow{2}{*}{Sample}
&  \multicolumn{2}{c|}{Initial} &  \multicolumn{2}{c}{BDT $>$ 0.92 } \\
&  $\Pe\Pe$ & $\mu\mu$ &  $\Pe\Pe$ & $\mu\mu$ \\
\hline
WW & \x62 $\pm$ 16   & 116 $\pm$ 22   & \NA & \NA  \\
WZ & 914 $\pm$ 38  & 2151 $\pm$ 63\x   & 1.6 $\pm$ 1.6    &1.8 $\pm$ 1.8    \\
ZZ & 522 $\pm$ 17   & 1324 $\pm$  29\x   & 1.8 $\pm$ 1.1    & 2.7 $\pm$ 1.3    \\
$\ttbar$ &  5363 $\pm$ 48\x  & 12938 $\pm$ 81\x\x   & 7.1 $\pm$ 1.9   & 7.1 $\pm$ 1.9    \\
single top quark & 269 $\pm$ 18   & 723 $\pm$ 31  & \NA  & \NA   \\
W + jets & 34 $\pm$ 5   & 36 $\pm$ 5   & \NA\  & \NA   \\ \hline
\dyzjj& 152750 $\pm$ 510\x\x   & 394640 $\pm$ 880\x\x   & 273 $\pm$ 20\x   & 493 $\pm$ 31\x   \\ \hline
Total backgrounds& 159890 $\pm$ 510\x\x  & 411890 $\pm$ 890\x\x   & 283 $\pm$ 29\x   &505 $\pm$ 43\x    \\
\ewkzjj signal & 2833 $\pm$ 10\x   & 6665 $\pm$ 16\x   & 194.9 $\pm$ 2.6\x\x     & 379.7 $\pm$ 3.9\x\x    \\
\hline
Data & 163640 & 422499 & 418 & 892  \\
\end{tabular}
\end{table*}

\subsection{Discriminating gluons from quarks}
\label{sec:gtag}

Jets in signal events are expected to originate from quarks,
while for background events it is more probable that jets are
initiated by a gluon.
A quark-gluon likelihood (QGL) discriminant~\cite{Chatrchyan:2013jya}
is evaluated for the two tagging jets with the intent of distinguishing
the nature of each jet.

The QGL discriminant exploits differences in the showering and fragmentation of gluons and quarks
by making use of the following internal jet composition observables: 
(i) the particle multiplicity of the jet, (ii) the  minor root-mean-square of distance between
the jet constituents
in the $\eta$-$\phi$ plane, and (iii) the  \pt
distribution function of the jet constituents, as defined in Ref.~\cite{CMS-PAS-JME-13-002}.

The variables are used as inputs to a likelihood
discriminant on gluon and quark jets constructed from simulated dijet events.
The performance of this QGL discriminant is
evaluated and validated using independent,
exclusive samples of $\PZ$ + jet and dijet data~\cite{CMS-PAS-JME-13-002}.
Comparisons of simulation predictions and data distributions
allow the derivation of corrections to the simulated QGL distributions
and define a systematic uncertainty band.

\subsection{Additional discriminating variables}
\label{sec:var}

An event balance variable, $R(\pt^{\text{hard}})$, is used to separate the signal from the background,
defined as
\begin{equation}
\ifthenelse{\boolean{cms@external}}
{ 
\begin{split} 
R(\pt^{\,\text{hard}}) & = 
\frac
{\abs{ \vec{p}_{\mathrm{T} \mathrm{j}_1}+\vec{p}_{\mathrm{T} \mathrm{j}_2}+\vec{p}_{\mathrm{T} \PZ}}}
{ \abs{\vec{p}_{\mathrm{T} \mathrm{j}_1}} +\abs{\vec{p}_{\mathrm{T} \mathrm{j}_2}} + \abs{\vec{p}_{\mathrm{T} \PZ}} } \\
& =
\frac
{ \abs{\vec{p}_{\mathrm{T}}^{\,\text{hard}}}}
{ \abs{\vec{p}_{\mathrm{T} \mathrm{j}_1}} +\abs{\vec{p}_{\mathrm{T} \mathrm{j}_2}} + \abs{\vec{p}_{\mathrm{T} \PZ}} } \; ,
\end{split}
}
{
R(\pt^{\,\text{hard}})=
\frac
{\abs{ \vec{p}_{\mathrm{T} \mathrm{j}_1}+\vec{p}_{\mathrm{T} \mathrm{j}_2}+\vec{p}_{\mathrm{T} \PZ}}}
{ \abs{\vec{p}_{\mathrm{T} \mathrm{j}_1}} +\abs{\vec{p}_{\mathrm{T} \mathrm{j}_2}} + \abs{\vec{p}_{\mathrm{T} \PZ}} }
=
\frac
{ \abs{\vec{p}_{\mathrm{T}}^{\,\text{hard}}}}
{ \abs{\vec{p}_{\mathrm{T} \mathrm{j}_1}} +\abs{\vec{p}_{\mathrm{T} \mathrm{j}_2}} + \abs{\vec{p}_{\mathrm{T} \PZ}} } \; ,
}
\end{equation}
where $\vec{p}_{\mathrm{T} \mathrm{j}_1}$, $\vec{p}_{\mathrm{T} \mathrm{j}_2}$ and
$\vec{p}_{\mathrm{T} \PZ}$ are, respectively, the transverse momenta of the two tagging jets and of the Z boson,
and the numerator is the estimator of the \pt for the
hard process, \ie $\pt^{\,\text{hard}}$.

Angular variables useful for signal discrimination include  
the difference between the rapidity of the
$\PZ$ boson $y_{\PZ}$ and the average rapidity of the two tagging jets, \ie
\begin{equation}
y^*=y_{\PZ}-\frac{1}{2}(y_{\mathrm{j}_1}+y_{\mathrm{j}_2}),
\end{equation}
and the $z^*$ Zeppenfeld variable~\cite{Schissler:2013nga}
defined as 
\begin{equation}
z^*=\frac{ y^* } { \Delta y_{\mathrm{jj}} }.
\end{equation}

\begin{figure*}[htp]
\centering
\includegraphics[width=\cmsFigWidth]{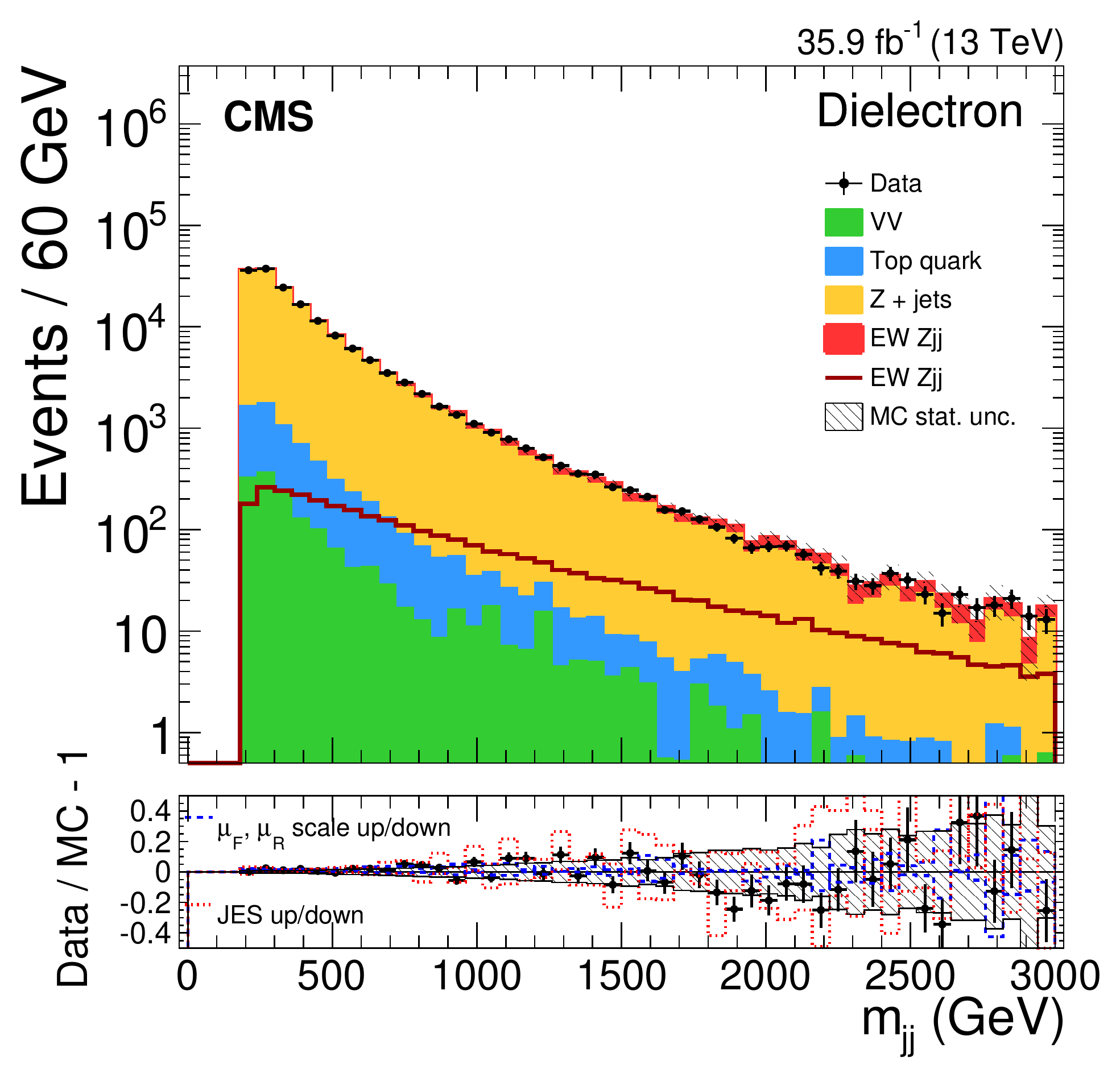} \hfil
\includegraphics[width=\cmsFigWidth]{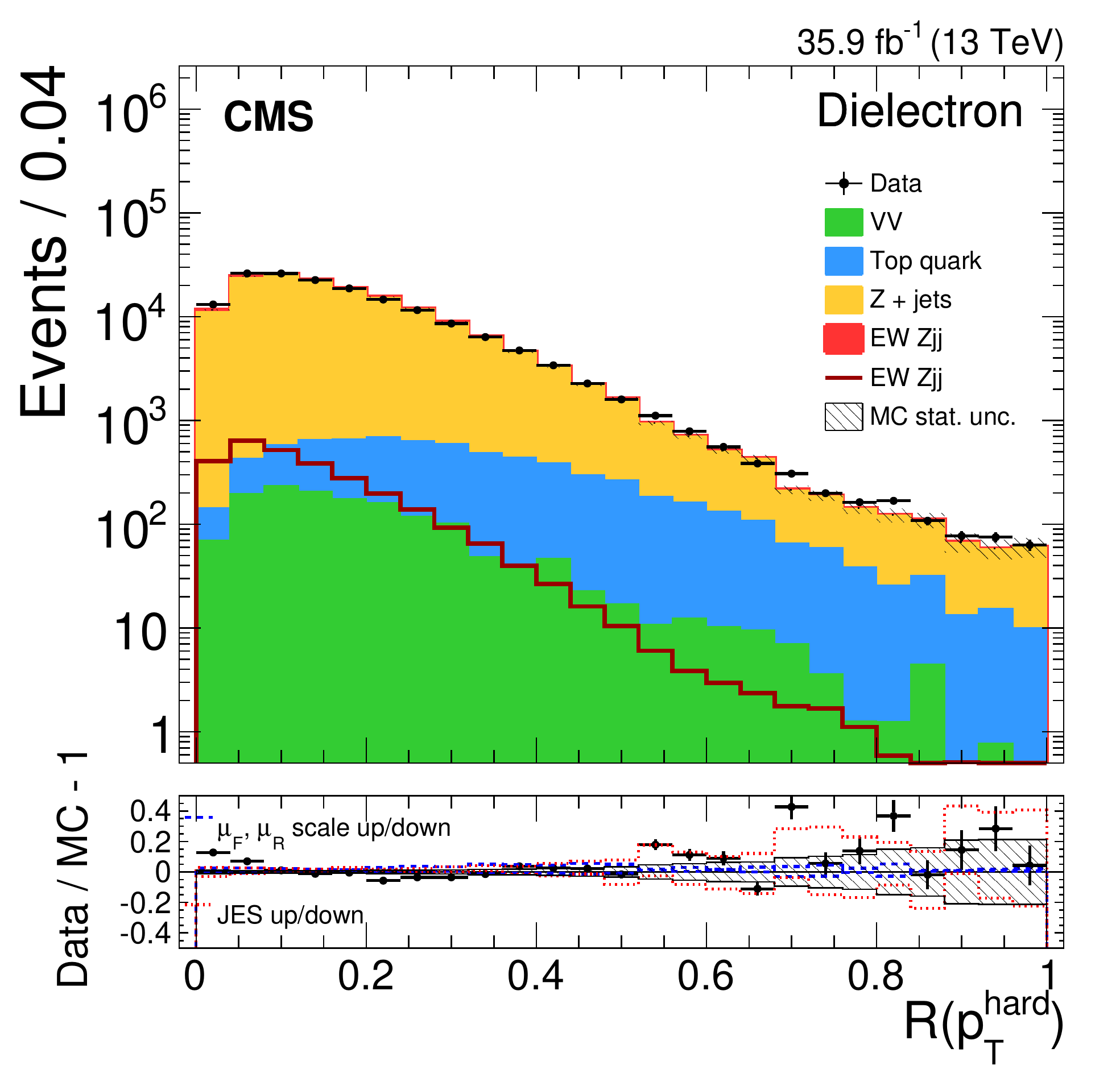} \\
\includegraphics[width=\cmsFigWidth]{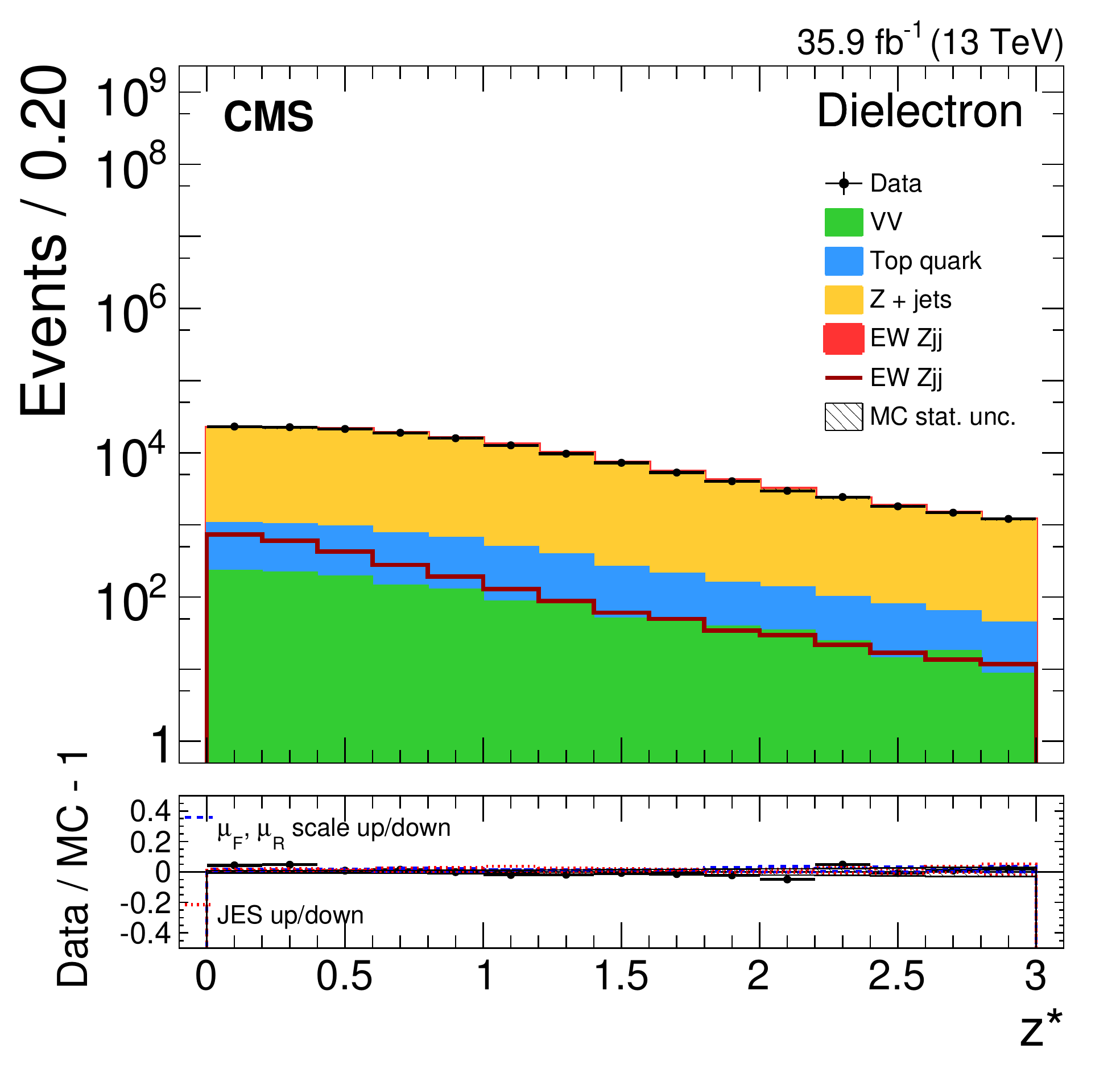}
\caption{
Data and simulated event distributions for the dielectron event selection:
$m_{\mathrm{jj}}$ (top left),
$R(\pt^{\,\text{hard}})$ (top right), and
$z^*$ (bottom).
The contributions from the different 
background sources and the signal are shown stacked, with data points superimposed.
The expected signal-only contribution is also shown as an unfilled histogram. 
The lower panels show
the relative difference between the data and expectations,
as well as the uncertainty
envelopes for JES and $\mu_{\rm F,R}$ scale uncertainties.
\label{fig:pre1_amc_el}
}
\end{figure*}

\begin{figure*}[htp]
\centering
\includegraphics[width=\cmsFigWidth]{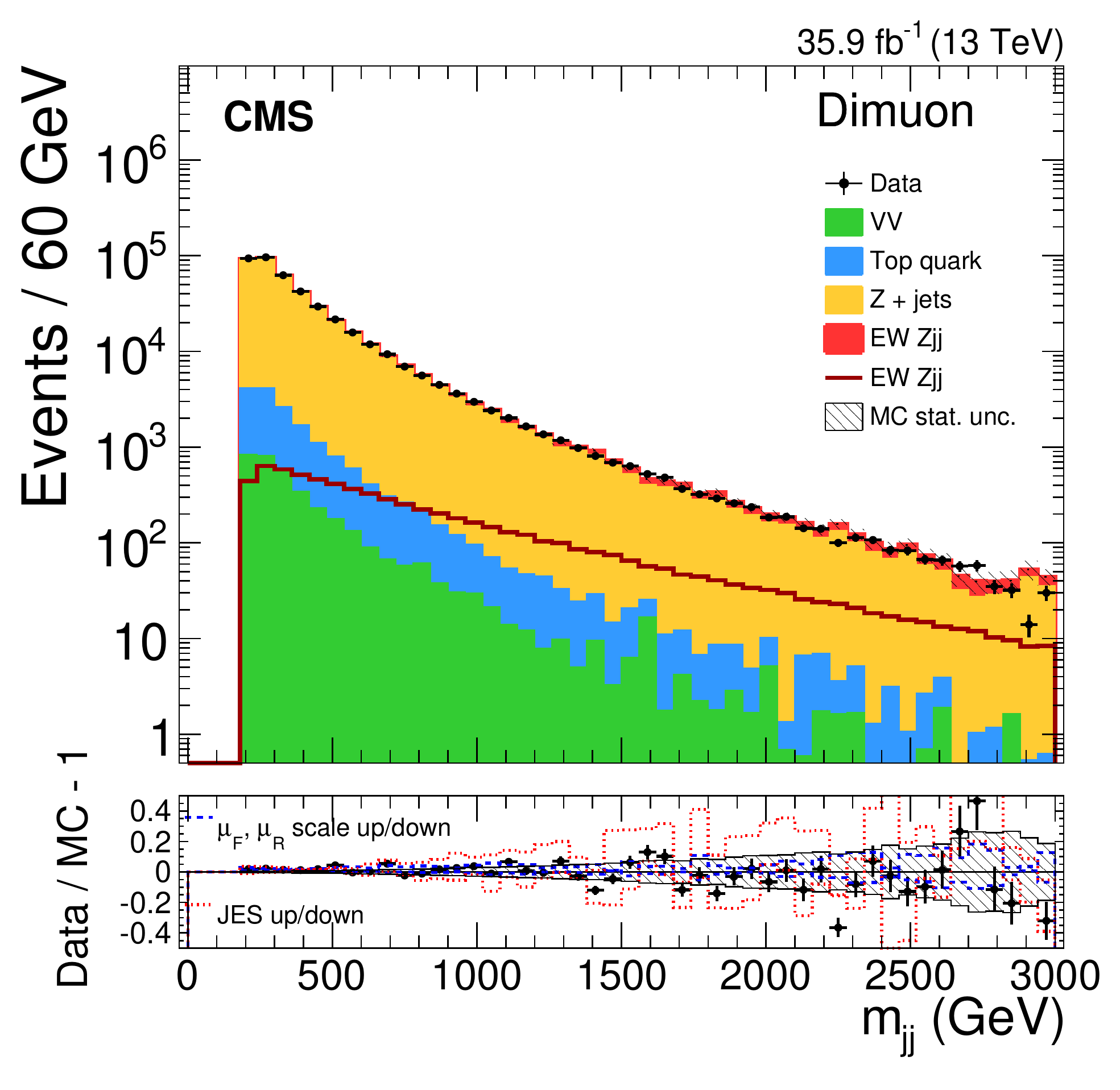} \hfil
\includegraphics[width=\cmsFigWidth]{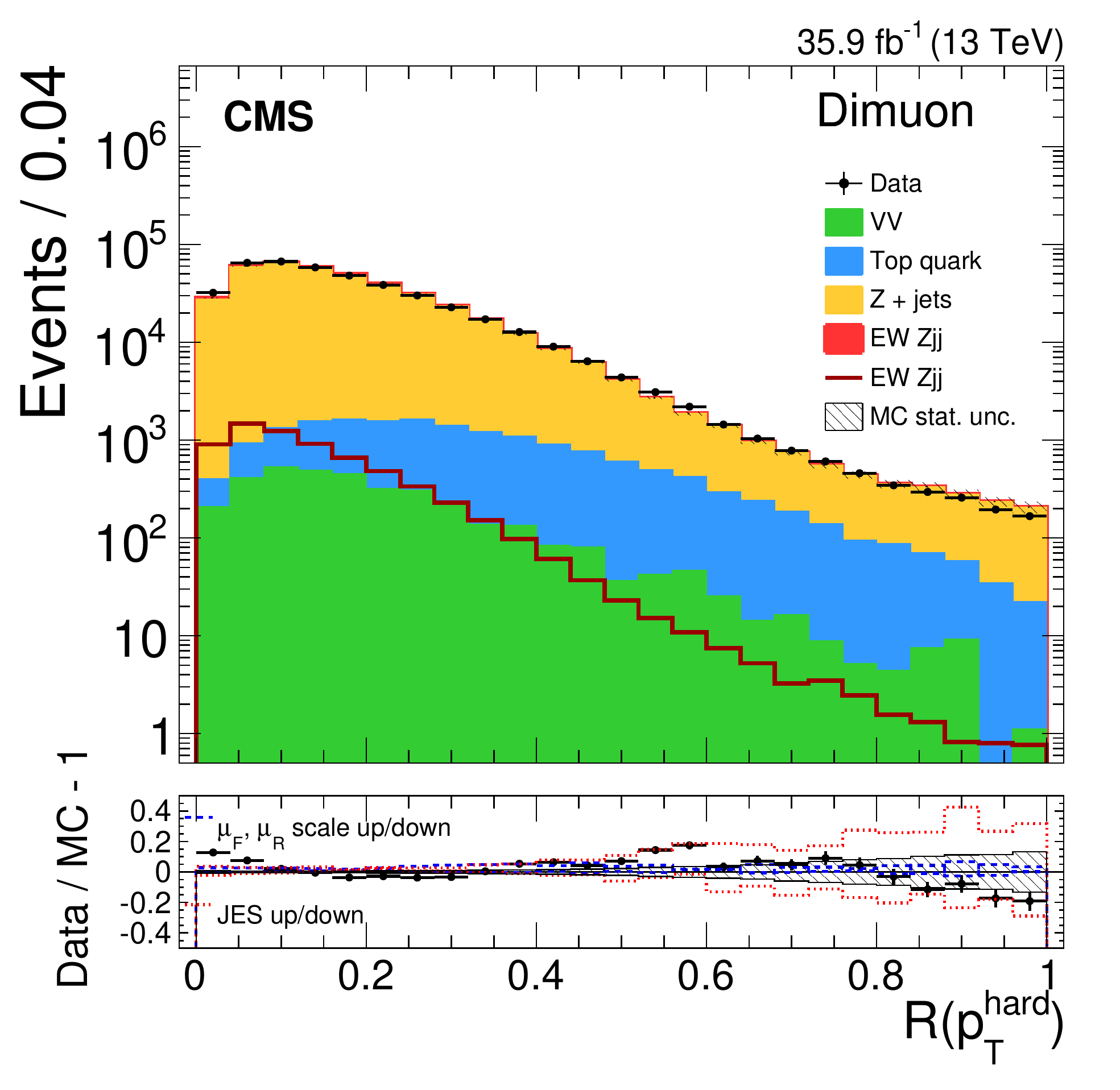} \\
\includegraphics[width=\cmsFigWidth]{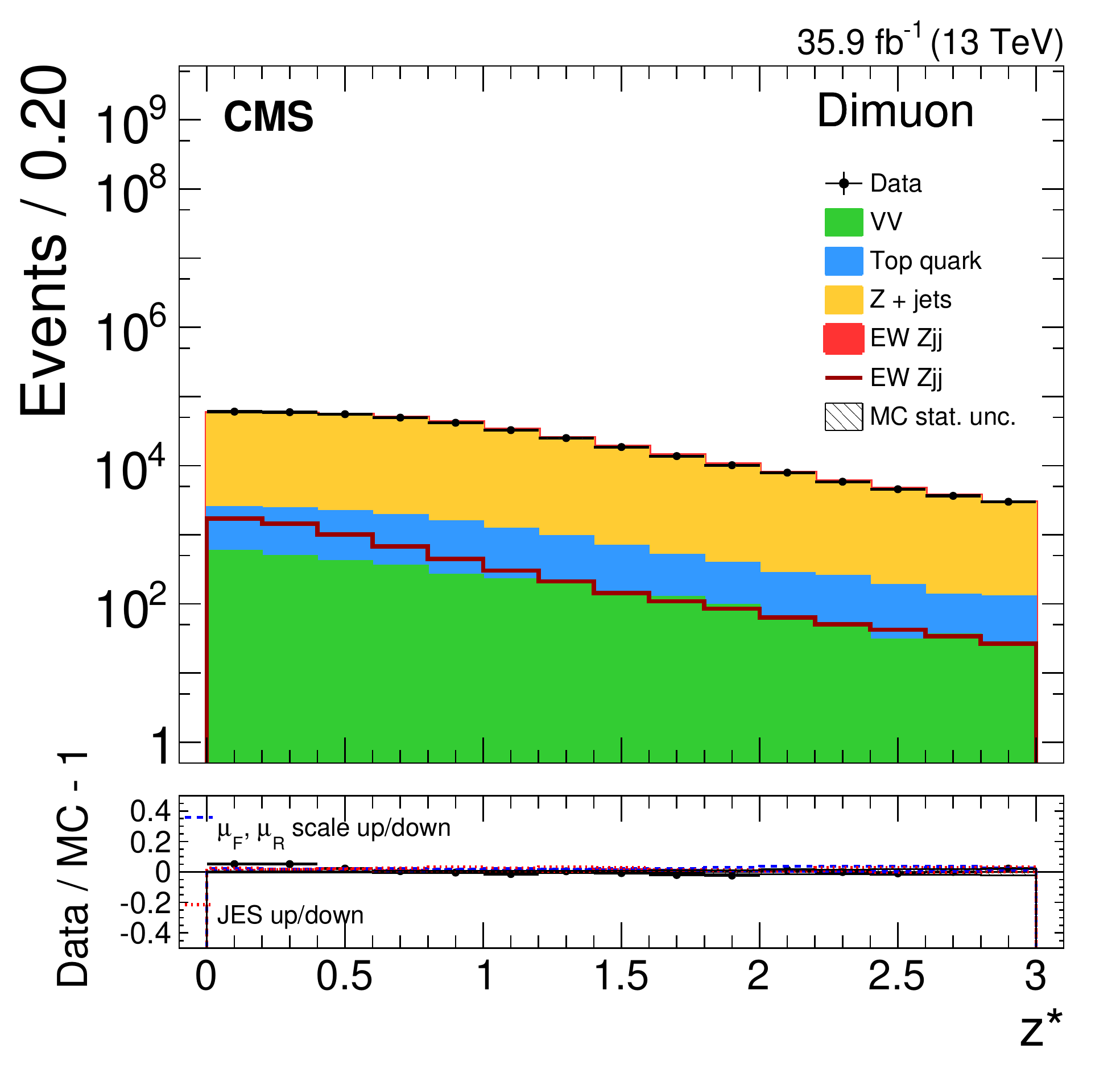}
\caption{
Data and simulated event distributions for the dimuon event selection:
$m_{\mathrm{jj}}$ (top left),
$R(\pt^{\,\text{hard}})$ (top right), and
$z^*$ (bottom).
The contributions from the different 
background sources and the signal are shown stacked, with data points superimposed.
The expected signal-only contribution is also shown as an unfilled histogram. 
The lower panels show
the relative difference between the data and expectations
as well as the uncertainty
envelopes for JES and $\mu_{\rm F,R}$  scale uncertainties.
\label{fig:pre1_amc_mu}
}
\end{figure*}

\begin{figure*}[htp]
\centering
\includegraphics[width=\cmsFigWidth]{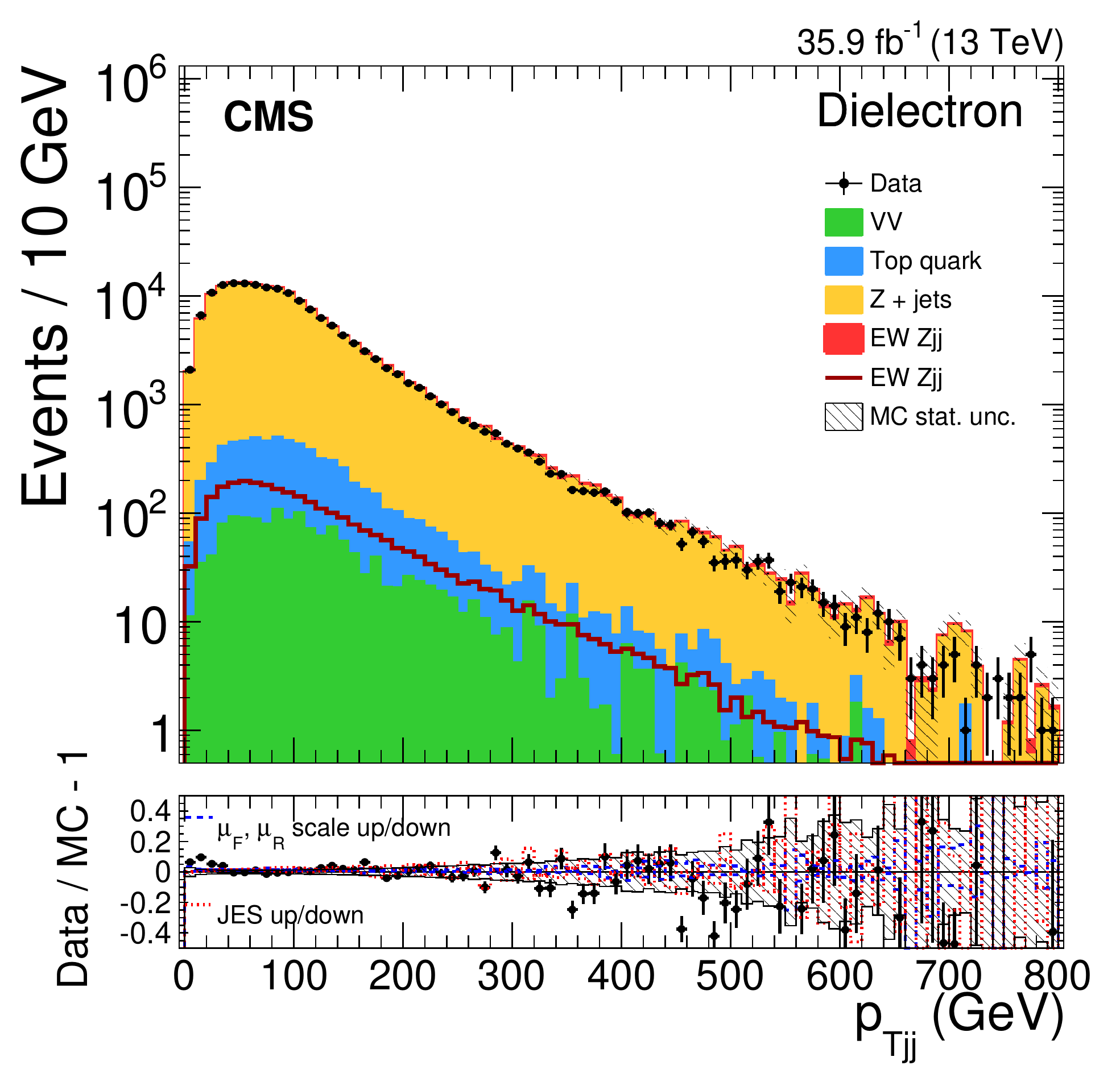} \hfil
\includegraphics[width=\cmsFigWidth]{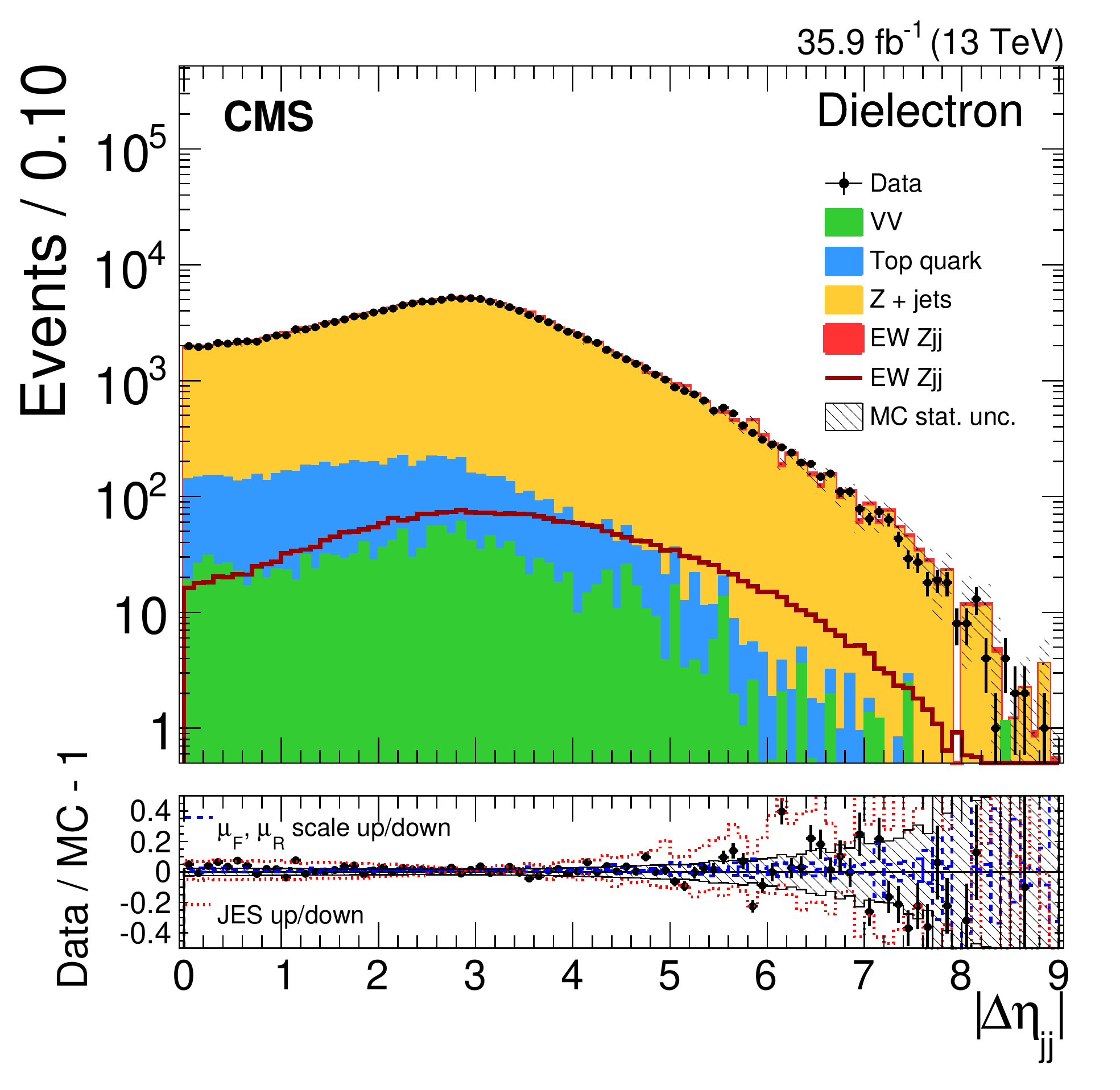} \\
\includegraphics[width=\cmsFigWidth]{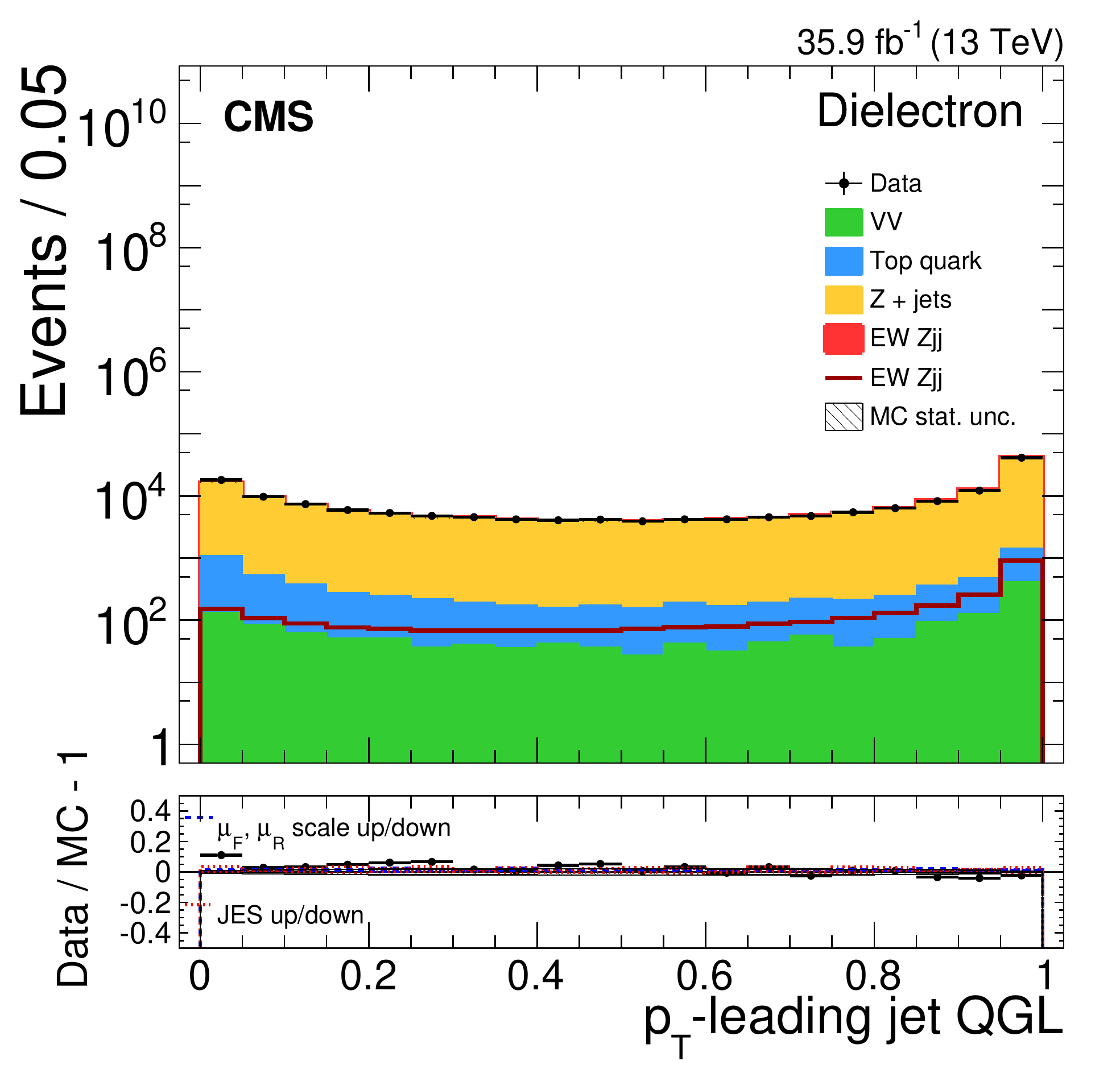} \hfil
\includegraphics[width=\cmsFigWidth]{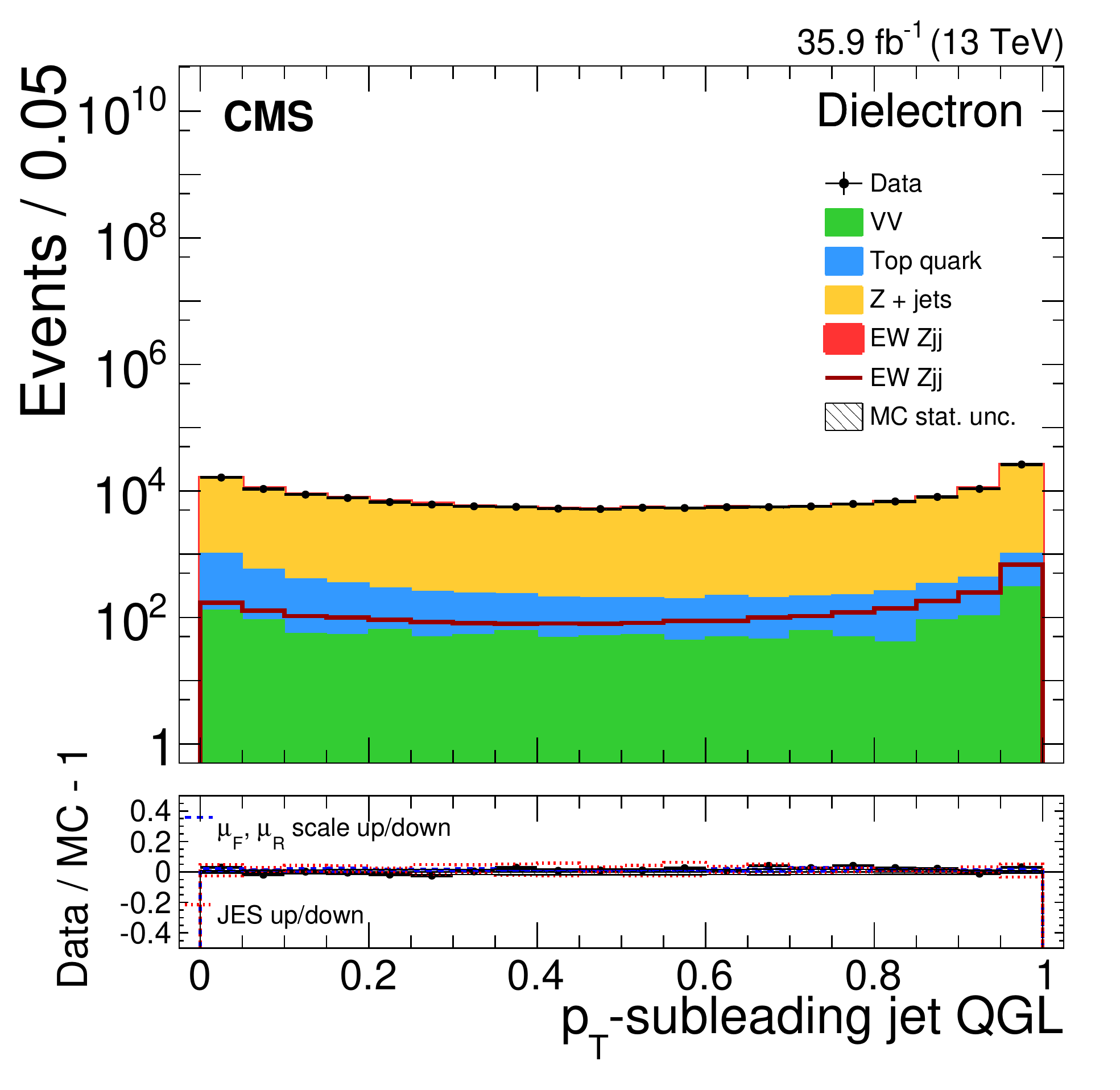}
\caption{
Data and simulated event distributions for the dielectron event selection:
dijet system transverse momentum (top left),
dijet pseudorapidity opening (top right), 
\pt-leading jet QGL (bottom left), and \pt-subleading jet QGL (bottom right).
The contributions from the different 
background sources and the signal are shown stacked, with data points superimposed.
The expected signal-only contribution is also shown as an unfilled histogram. 
The lower panels show
the relative difference between the data and expectations,
as well as the uncertainty
envelopes for JES and $\mu_{\rm F,R}$  scale uncertainties.
\label{fig:pre2_amc_el}
}
\end{figure*}

\begin{figure*}[htp]
\centering
\includegraphics[width=\cmsFigWidth]{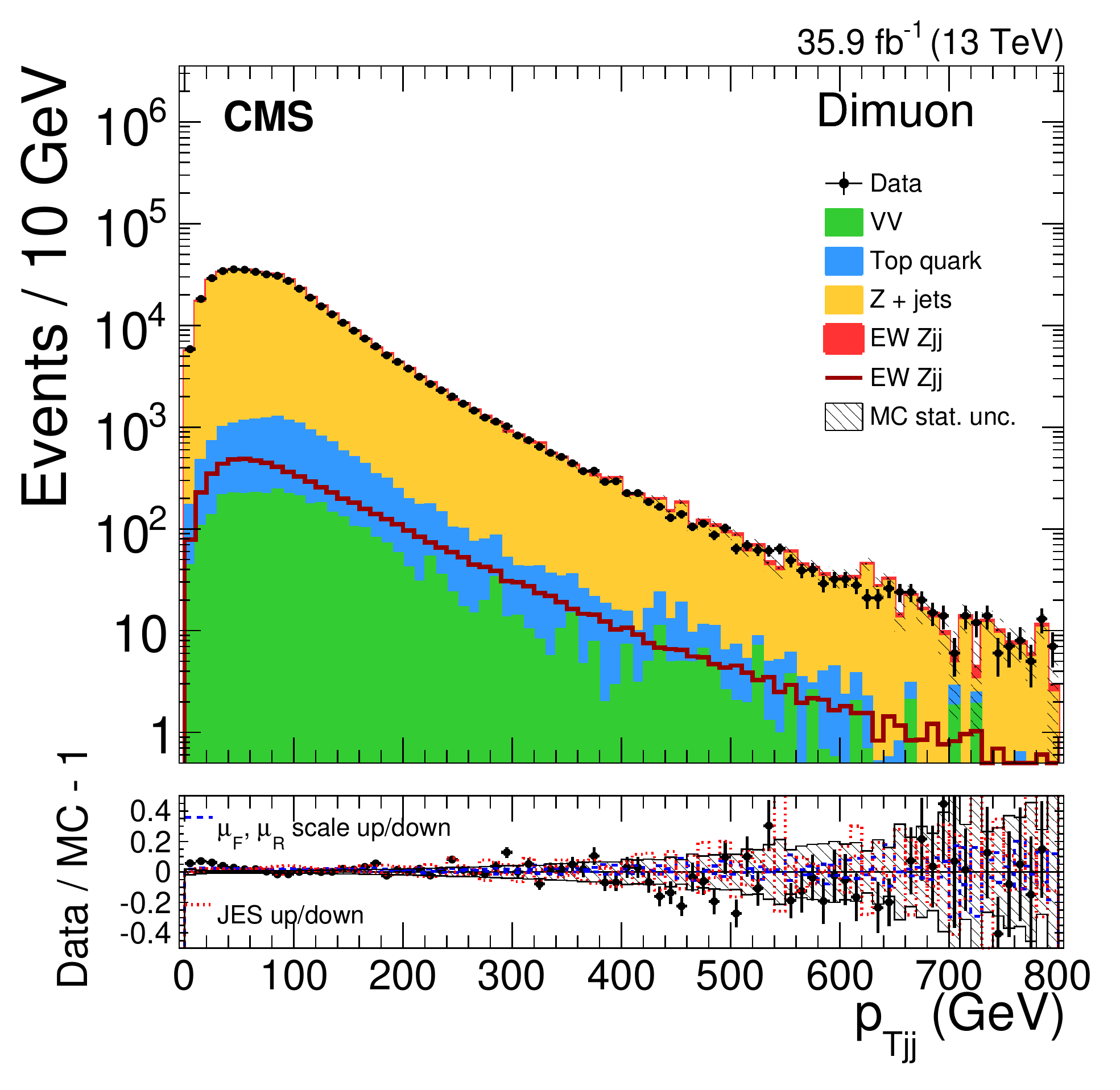} \hfil
\includegraphics[width=\cmsFigWidth]{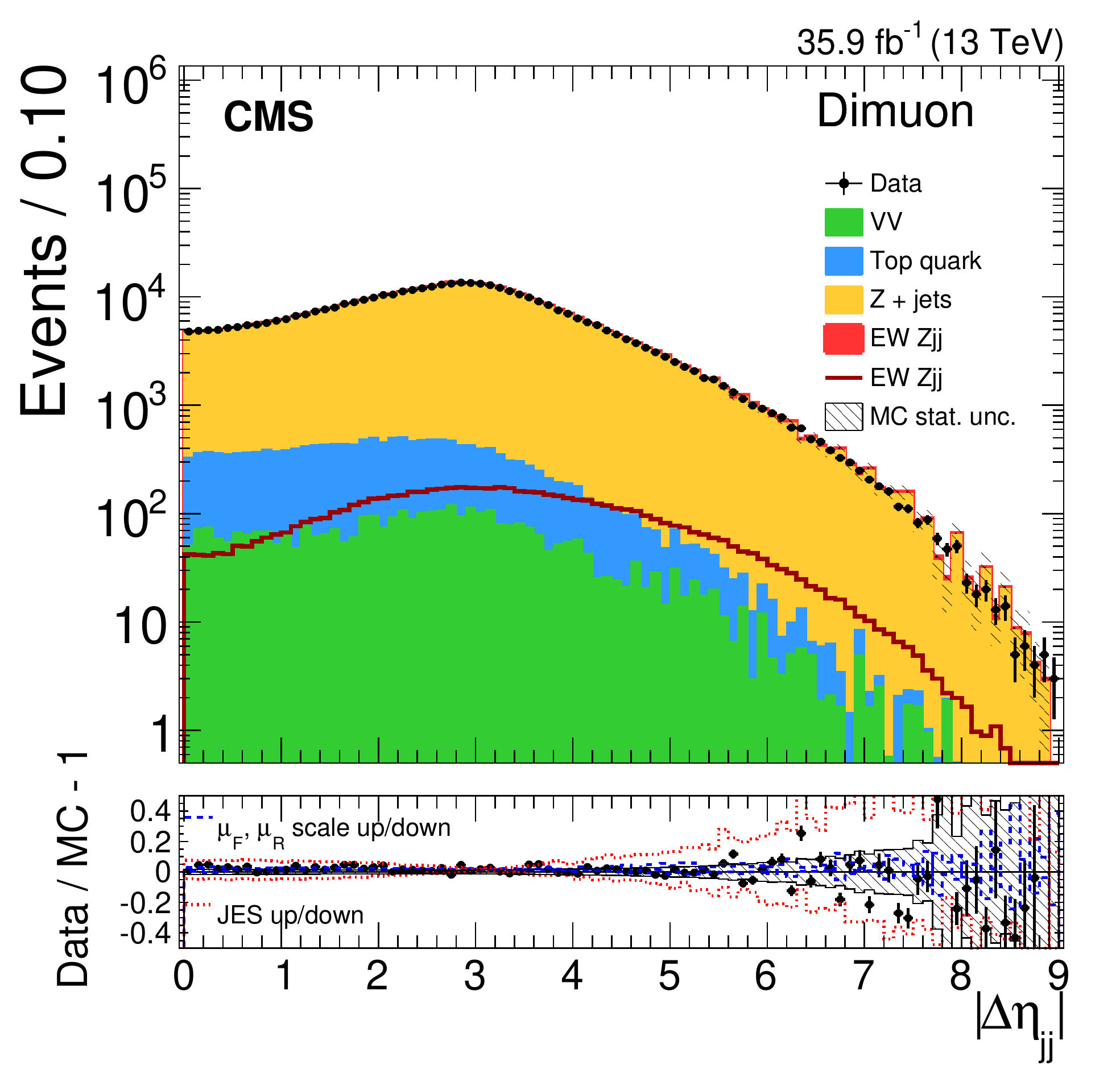} \\
\includegraphics[width=\cmsFigWidth]{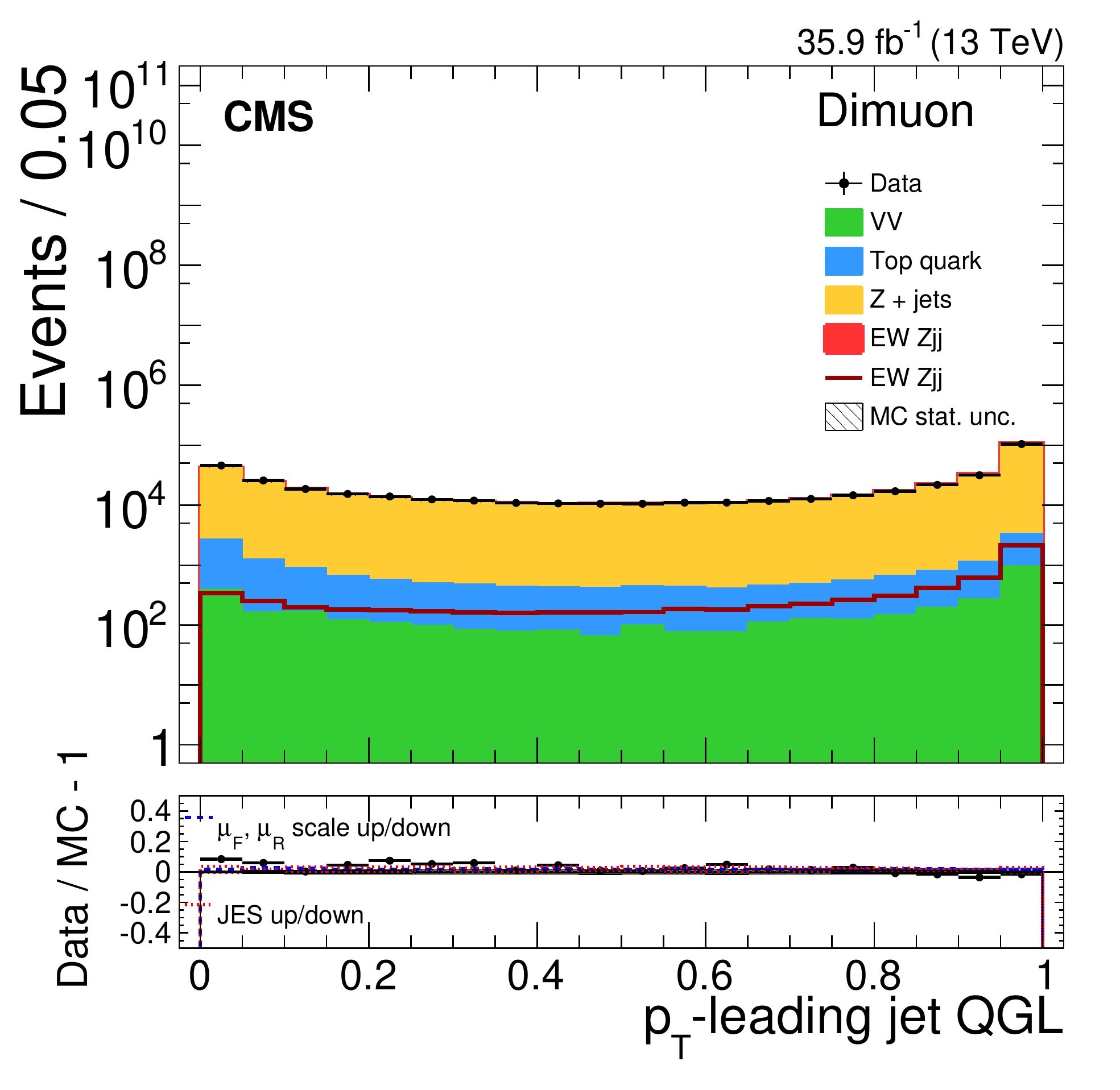} \hfil
\includegraphics[width=\cmsFigWidth]{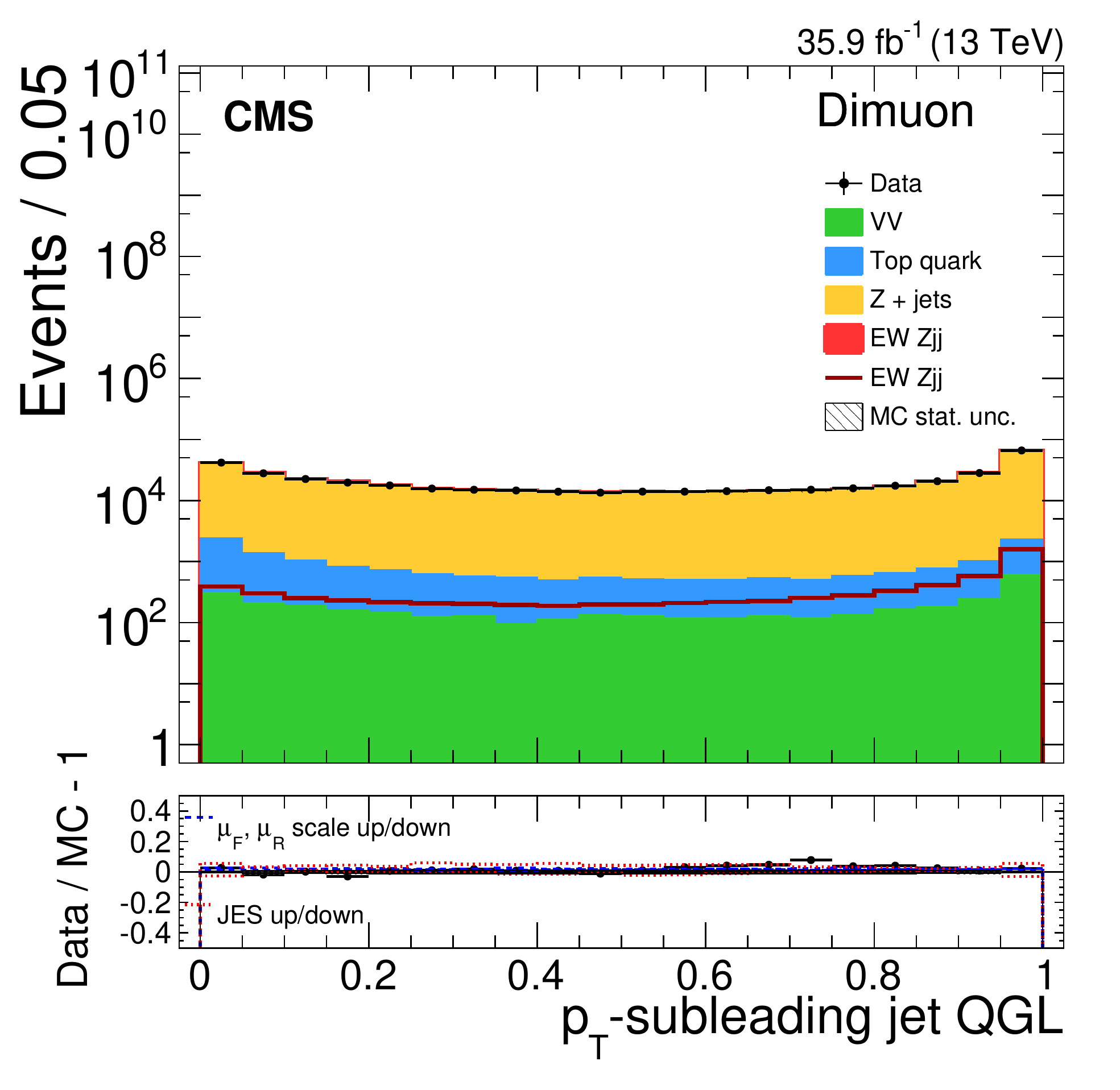}
\caption{
Data and simulated event distributions for the dimuon event selection:
dijet system transverse momentum (top left),
dijet pseudorapidity opening (top right), 
\pt-leading jet QGL (bottom left), and \pt-subleading jet QGL (bottom right).
The contributions from the different 
background sources and the signal are shown stacked, with data points superimposed.
The expected signal-only contribution is also shown as an unfilled histogram. 
The lower panels show
the relative difference between the data and expectations,
as well as the uncertainty
envelopes for JES and $\mu_{\rm F,R}$ scale uncertainties.
\label{fig:pre2_amc_mu}}
\end{figure*}

The distributions for data and simulated samples 
of the $m_{\mathrm{jj}}$, $R(\pt^{\,\text{hard}})$ and
$z^*$ variables, after the initial selection, are shown in Figs.~\ref{fig:pre1_amc_el} and \ref{fig:pre1_amc_mu},
for the dielectron and dimuon channels, respectively.
The distributions for data and simulated samples 
of the dijet transverse momentum 
($\ptjj$), pseudorapidity separation ($\Delta\eta_{\mathrm{jj}}$), and
of the QGL output values of each jet, after the initial selection,
are shown in Figs.~\ref{fig:pre2_amc_el} and \ref{fig:pre2_amc_mu},
respectively, for the
dielectron and dimuon channels. 
Good agreement between the data and the MC expectations is observed in both channels.
In the lower panels of these plots the experimental uncertainties in the jet energy scales (JES)
(dotted envelope) and the uncertainties due to the choice of QCD factorisation and normalization scales
defined in Sec.~\ref{sec:systunc} (dashed envelope) are shown.

\section{Signal discriminants and extraction procedure}
\label{sec:sigdisc}
The \ewkzjj\ signal is characterized
by a  large separation in pseudorapidity between the tagging jets,
due to the small scattering-angle of the two initial partons.
Because of both the topological configuration and the large energy
of the outgoing partons, $m_{\mathrm{jj}}$ is also
expected to be large.
The evolution of $\Delta\eta_{\mathrm{jj}}$ with $m_{\mathrm{jj}}$ is expected
to be different for signal and background events, and therefore
these characteristics are expected to yield a high separation power
between the \ewkzjj\ and the \dyzjj\ productions.
In addition, in signal events
it is expected that the Z boson candidate is produced 
centrally 
in the rapidity region defined by the two tagging jets
and that the \Zjj\ 
system is approximately balanced in the transverse plane.
As a consequence signal events are expected to yield lower values
of both $z^*$
and $\pt^{\,\text{hard}}$ than the DY background.
Other variables that are used to enhance the signal-to-background separation
are related to the kinematics of the event
(\pt, rapidity, and distance between the jets and/or the $\PZ$ boson)
or to the properties of the jets that are expected to be initiated by
quarks.
The variables that are used in the 
multivariate analysis are:
(i)  $m_{\rm jj}$;
(ii) $\Delta\eta_{\rm jj}$;
(iii) the dijet transverse momentum \ptjj;
(iv) the QGL values of the two tagging jets;
(v) $R\left(\pt^{\,\text{hard}}\right)$ and $z^*$.

The output of the discriminator is built by training a boosted decision tree
(BDT) from the \textsc{tmva} package~\cite{Hocker:2007ht}
to achieve an optimal separation between the
\ewkzjj\ and \dyzjj\ processes, independently in the dielectron and dimuon channels.

In order to improve the sensitivity for the extraction of the signal 
component, the transformation that originally projects the BDT output 
value in the [$-1$,$+1$] interval is changed into  ${\rm BDT'} = \tanh^{-1}(({\rm BDT}+1)/2)$.
This allows the purest signal region of the BDT output to be better 
sampled while keeping an equal-width binning of the BDT variable.

\begin{figure*}[htp] {
\centering
\includegraphics[width=\cmsFigWidth]{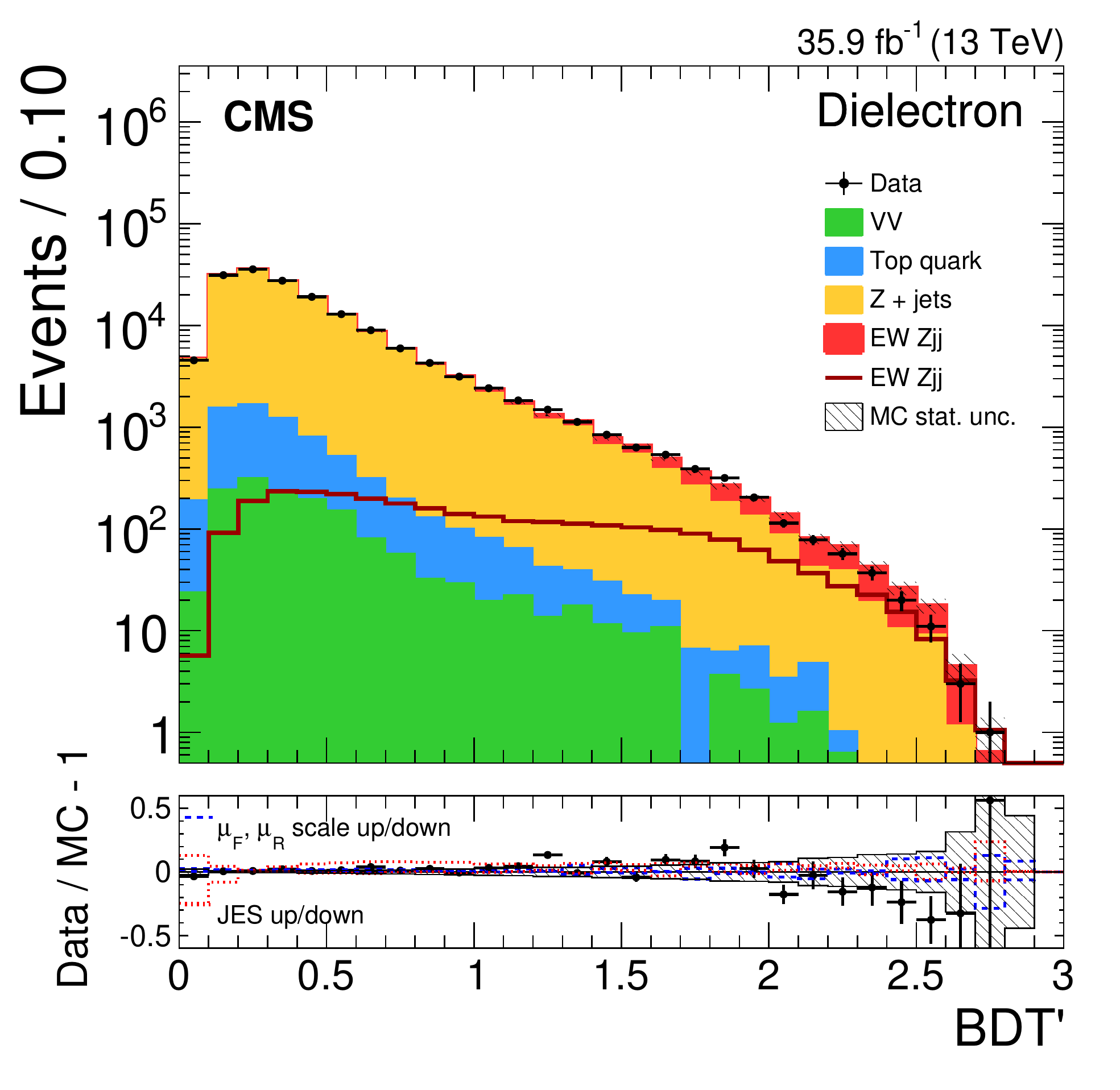} \hfil
\includegraphics[width=\cmsFigWidth]{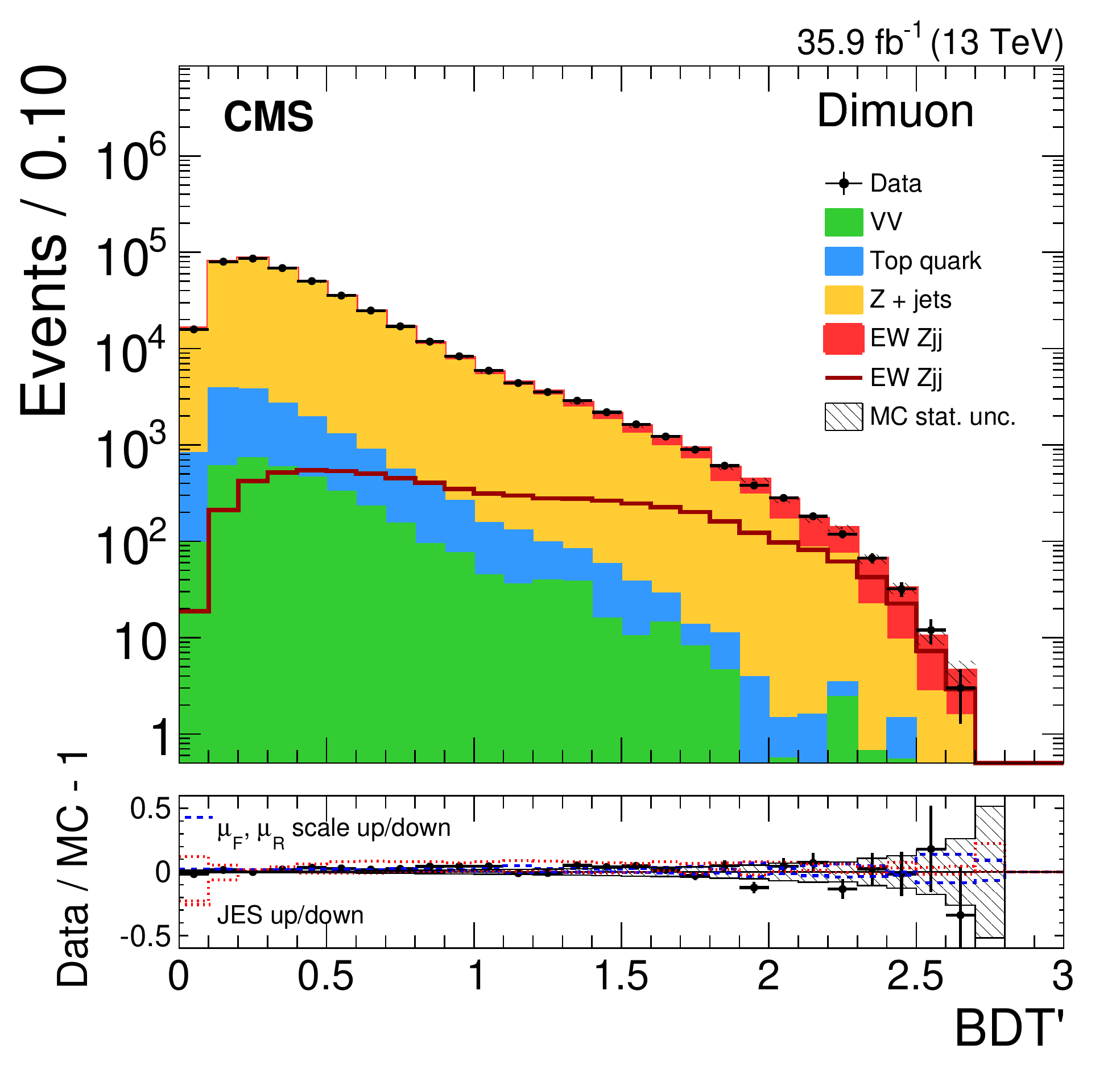}
\caption{
Distributions for transformed BDT discriminants in dielectron (left) and dimuon
(right) events.
The contributions from the different 
background sources and the signal are shown stacked, with data points superimposed.
The expected signal-only contribution is also shown as an unfilled histogram. 
The lower panels show
the relative difference between the data and expectations,
as well as the uncertainty
envelopes for JES and $\mu_{\rm F,R}$ scale uncertainties.
}
\label{fig:bdt}
}
\end{figure*}

Figure~\ref{fig:bdt} shows
the distributions of the discriminants for the two leptonic channels.
Good overall agreement between simulation and data is observed in all distributions,
and the signal presence is visible at high BDT' values.

{\tolerance=1200
A binned maximum likelihood calculation, which is used to fit simultaneously 
the strength modifiers for the \ewkzjj\ and \dyzjj\ processes,
$\mu = \sigma({\mathrm{EW}\,\PZ\mathrm{jj}}) / \sigma_\mathrm{LO}({\mathrm{EW}\,\ell\ell\mathrm{jj}})$
and
$\upsilon = \sigma({\mathrm{DY}})/\sigma_\text{th}({\mathrm{DY}})$,
is built from the expected
rates for each process.
Nuisance parameters are added to modify the
expected rates and shapes according to the estimate of the systematic
uncertainties affecting the measurement.
\par}

The interference between the \ewkzjj\ and
\dyzjj\ processes is included in the fit procedure,
and its strength scales as $\sqrt{\mu\upsilon}$. 
The interference model is derived from
the \MGvATNLO simulation described in Section~\ref{sec:simulation}.

The parameters of the model ($\mu$ and $\upsilon$) are determined by
maximizing the likelihood.
The statistical methodology follows the one used in other CMS
analyses~\cite{Chatrchyan:2012ufa}
using the asymptotic formulas~\cite{Cowan:2010js}.
In this procedure the systematic uncertainties affecting the
measurement of the signal and background strengths ($\mu$ and $\upsilon$)
are constrained with log-normal probability distributions.

\section{Systematic uncertainties}
\label{sec:systunc}

{\tolerance=1200
The main systematic uncertainties affecting the measurement
are classified into experimental and theoretical sources.
Some uncertainties affect only 
normalizations, while others affect both the normalization and shape
of the BDT output distribution. 
\par}

\subsection{Experimental uncertainties}
\label{subsec:expunc}

The following experimental uncertainties are considered.

{\tolerance=1200

\begin{description}

\item[Integrated luminosity] --- A 2.5\% uncertainty is assigned to the value of
the integrated luminosity~\cite{CMS-PAS-LUM-17-001}.

\item[Trigger and selection efficiencies] ---
Uncertainties in the efficiency corrections based on control samples in data for
the leptonic trigger and 
offline selections amount to a total of 2--3\%,
depending on the lepton $\pt$ and $\eta$ 
for both the ee and $\mu\mu$ channels. These uncertainties are
estimated by comparing the lepton efficiencies expected in
simulation and measured in data with
a tag-and-probe method~\cite{Khachatryan:2010xn}.

\item[Jet energy scale and resolution] --- The energy of the jets enters
at the selection level and in the
computation of the kinematic variables used to calculate the discriminants.
Therefore the uncertainty in the JES affects both the
expected event yields and the final shapes.
The effect of the JES uncertainty is studied by
scaling up and down the reconstructed jet energy
by \pt- and $\eta$-dependent scale factors~\cite{Chatrchyan:2011ds}.
An analogous approach is used for the JER.
The final impact on the signal strength uncertainty
amounts to about 3\% for JES and 2\% for JER.

\item[QGL discriminator] --- The uncertainty in the performance of the
QGL discriminator is measured
using independent $\PZ$ + jet and dijet data~\cite{CMS-PAS-JME-13-002}.
Shape variations corresponding to the full data versus simulation differences
are implemented. The variations are of the order of 10\% for lower QGL output values,
corresponding to gluon-like jets, and of the order of 5\% for larger QGL output values,
corresponding to quark-like jets.
The final impact on the signal strength uncertainty
amounts to about 1\%.

\item[Pileup] --- Pileup can affect the identification and
isolation of the leptons or the corrected energy of the jets. When
jet clustering is performed, pileup can
induce a distortion of the reconstructed dijet system because of the contamination
from tracks and calorimetric deposits.
This uncertainty is evaluated by generating
alternative distributions of the 
number of pileup interactions, corresponding 
to a 5\% uncertainty in the total
inelastic pp cross section at $\sqrt{s}=13\TeV$.

\item[Limited number of simulated events] --- For each signal and background simulation, 
shape variations for the distributions are created by shifting each bin
content up or down by its statistical uncertainty.
This generates alternatives to the nominal shapes that are used to describe the uncertainty 
from the limited number of simulated events.
Depending on the BDT output bin, the impact on the signal strength uncertainty
can be up to 3\%.

\end{description}

\par}

\subsection{Theoretical uncertainties}
\label{subsec:thunc}

The following theoretical uncertainties are considered in the analysis.

{\tolerance=1200

\begin{description}

\item[PDF] --- The PDF uncertainties are evaluated
by comparing the nominal distributions to those obtained when using the alternative PDFs of the NNPDF set,
including $\alpha_\mathrm{S}$ variations.
The final impact on the signal strength uncertainty
is less than 1\%.

\item[Factorization and renormalization scales] ---
To account for theoretical uncertainties, signal and background shape variations
are built by changing the values of $\mu_{\rm F}$ and $\mu_{\rm R}$ from their defaults
by factors of 2 or 1/2  in the ME calculation, simultaneously for
$\mu_{\rm F}$ and $\mu_{\rm R}$, but independently for each simulated sample. 
The final impact on the signal strength uncertainty
amounts to 6\% and 4\% respectively for the signal and background variations.

\item[Normalization of top quark and diboson backgrounds] --- \ifthenelse{\boolean{cms@external}}{\\}{}Diboson and top quark
production  processes are modelled with MC simulations.
An uncertainty in the normalization of these backgrounds
is assigned based on the PDF and $\mu_{\rm F}$, $\mu_{\rm R}$
uncertainties, following calculations
in Refs.~\cite{Campbell:2010ff,Czakon:2013goa,Kidonakis:2012db}.
The final impact on the signal strength uncertainty
amounts to less than 1\%.

\item[Interference between \ifthenelse{\boolean{cms@external}}{\ewkzjj}{\boldmath{\ewkzjj}} and \ifthenelse{\boolean{cms@external}}{\dyzjj}{\boldmath{\dyzjj}}] --- An overall normalization
uncertainty and a shape uncertainty
are  assigned to the interference term in the fit, based on an envelope of prediction with
different $\mu_{\rm F}$, $\mu_{\rm R}$ scales. 
The final impact on the signal strength uncertainty
amounts to 2--3\%.

\item[Parton showering model] --- The uncertainty in the signal PS
model and the event tune is assessed as the full difference
of the acceptance and shape predictions using \PYTHIA and \HERWIGpp. 
The final impact on the signal strength uncertainty
amounts to about 4\%.

\end{description}

\par}

The largest sources of experimental uncertainty come from the JES and the
limited statistics of simulated events;
the largest source of theoretical uncertainty comes from the
$\mu_{\rm F}$, $\mu_{\rm R}$ scale uncertainties.

\section{Measurement of the EW Zjj production cross section}
\label{sec:results}
The signal strength, defined 
for the $\ell\ell\mathrm{jj}$ final state
in the kinematic region described in Sec.~\ref{sec:simulation},
is extracted from the fit to the BDT output distribution
as discussed in Section~\ref{sec:sigdisc}.

In the dielectron channel, the signal strength is measured to be 
\begin{equation*}
\ifthenelse{\boolean{cms@external}}
{ 
\begin{split} 
\mu & =0.96 \pm 0.06\stat \pm0.13\syst \\ & =0.96\pm 0.14\,\text{(total)},
\end{split}
}
{
\mu=0.96 \pm 0.06\stat \pm0.13\syst=0.96\pm 0.14\,\text{(total)},
}
\end{equation*}
corresponding to a measured signal cross section
\begin{equation*}
\ifthenelse{\boolean{cms@external}}
{ 
\begin{split} 
\sigma({\mathrm{EW}~\ell\ell\mathrm{jj}}) & =
521 \pm 34\stat \pm 68\syst\unit{fb} \\ & =521\pm 76 \,\text{(total)}\unit{fb}.
\end{split}
}
{
\sigma({\mathrm{EW}~\ell\ell\mathrm{jj}})=
521 \pm 34\stat \pm 68\syst\unit{fb}=521\pm 76 \,\text{(total)}\unit{fb}.
}
\end{equation*}

In the dimuon channel, the signal strength is measured to be 
\begin{equation*}
\ifthenelse{\boolean{cms@external}}
{ 
\begin{split} 
\mu & =0.97\pm 0.04\stat \pm0.11\syst \\ & =0.97\pm 0.12 \,\text{(total)},
\end{split}
}
{
\mu =0.97\pm 0.04\stat \pm0.11\syst =0.97\pm 0.12 \,\text{(total)},
}
\end{equation*}
corresponding to a measured signal cross section
\begin{equation*}
\ifthenelse{\boolean{cms@external}}
{ 
\begin{split} 
\sigma({\mathrm{EW}~\ell\ell\mathrm{jj}}) & =
524 \pm 23\stat \pm 61\syst\unit{fb} \\ & =524\pm 65\,\text{(total)}\unit{fb}.
\end{split}
}
{
\sigma({\mathrm{EW}~\ell\ell\mathrm{jj}}) =
524 \pm 23\stat \pm 61\syst\unit{fb}=524\pm 65\,\text{(total)}\unit{fb}.
}
\end{equation*}

The results obtained for the different dilepton channels are compatible with each other, and in 
agreement with the SM predictions.

{\tolerance=600
From the combined fit of the two channels, the signal strength is measured to be 
\begin{equation*}
\ifthenelse{\boolean{cms@external}}
{ 
\begin{split} 
\mu & =0.98\pm 0.04\stat \pm0.10\syst \\ & =0.98\pm 0.11\,\text{(total)},
\end{split}
}
{
\mu=0.98\pm 0.04\stat \pm0.10\syst=0.98\pm 0.11\,\text{(total)},
}
\end{equation*}
corresponding to a measured signal cross section
\begin{equation*}
\ifthenelse{\boolean{cms@external}}
{ 
\begin{split} 
\sigma({\mathrm{EW}~\ell\ell\mathrm{jj}}) & =
534 \pm 20\stat \pm 57\syst\unit{fb} \\ & =534\pm 60\,\text{(total)}\unit{fb},
\end{split}
}
{
\sigma({\mathrm{EW}~\ell\ell\mathrm{jj}})=
534 \pm 20\stat \pm 57\syst\unit{fb}=534\pm 60\,\text{(total)}\unit{fb},
}
\end{equation*}
in agreement with the SM prediction
$\sigma_\mathrm{LO}(\mathrm{EW}\,\ell\ell\mathrm{jj})=543\pm 24\unit{fb}$.
In the combined fit, the DY strength is $\upsilon = 0.988 \pm 0.031$.
Using the statistical methodology described in Ref.~\cite{Cowan:2010js},
the background-only hypotheses in the dielectron, dimuon, and combined
channels are all excluded with significance well above 5$\sigma$.
\par}

\section{Limits on anomalous gauge couplings}
\label{sec:atgc}

In the framework of EFT, new physics can be 
described as an infinite series of new interaction terms organized as an 
expansion in the mass dimension of the operators. 

In the EW sector of the SM, the first higher-dimensional 
operators containing bosons are six-dimensional~\cite{Degrande:2012wf}:
\begin{equation}\label{atgc:eq1}
\begin{split}
\mathcal{O}_{WWW} & = \frac{c_{WWW}}{\Lambda^2}W_{\mu\nu}W^{\nu\rho}W_{\rho}^{\mu},\\
\mathcal{O}_{W} & = \frac{c_{W}}{\Lambda^2}(D^{\mu}\Phi)^{\dagger}W_{\mu\nu}(D^{\nu}\Phi),\\
\mathcal{O}_{B} & = \frac{c_{B}}{\Lambda^2}(D^{\mu}\Phi)^{\dagger}B_{\mu\nu}(D^{\nu}\Phi),\\
\widetilde{\mathcal{O}}_{WWW} & = \frac{\widetilde{c}_{WWW}}{\Lambda^2}\widetilde{W}_{\mu\nu}W^{\nu\rho}W_{\rho}^{\mu},\\
\widetilde{\mathcal{O}}_{W} & = \frac{\widetilde{c}_{W}}{\Lambda^2}(D^{\mu}\Phi)^{\dagger}\widetilde{W}_{\mu\nu}(D^{\nu}\Phi),
\end{split}
\end{equation}
where, as is customary, group indices are suppressed and the mass scale
$\Lambda$ is factorized
from the coupling constants $c$. In Eq.~\eqref{atgc:eq1}, $W_{\mu\nu}$ 
is the SU(2) field strength, $B_{\mu\nu}$ is the U(1) field strength, $\Phi$ 
is the Higgs doublet, and operators with a tilde are the magnetic duals of the 
field strengths. The first three operators are charge and parity conserving, whereas the two 
last ones are not.
In this paper, models with operators that preserve charge conjugation and parity symmetries
can be included in the calculation either individually or in pairs.
With these assumptions,
the value of coupling constants divided by the mass scale $c/\Lambda^2$ are measured.

These operators have a rich phenomenology since they contribute to many  
multiboson scattering processes at tree level.
The operator $\mathcal{O}_{WWW}$ modifies 
vertices with 3 to 6 vector bosons, whereas the operators $\mathcal{O}_{W}$ and 
$\mathcal{O}_{B}$ modify both HVV vertices and vertices with 3 or 4 vector 
bosons. A more detailed description of the phenomenology of these operators 
can be found in Ref.~\cite{Degrande:2013yda}.
Modifications to the ZWW vertex are investigated in this case, since this modifies
the $ \Pp\Pp \to \PZ \mathrm{jj} $ cross section.

Previously, modifications to these vertices have been studied using
anomalous trilinear gauge couplings~\cite{pdg}.
The relationship between the dimension-6 operators in Eq.~\eqref{atgc:eq1}
and ATGCs can be found in Ref.~\cite{Degrande:2012wf}. 

\subsection{ATGC signal simulation}\label{atgc:sec2}

ATGC signal events are simulated at LO using  \MGvATNLO  with the \textsc{NNPDF3.0} 
PDF set for signal generation. Showering and hadronization 
of the events is performed with \PYTHIA
using the CUETP8M1 tune~\cite{Khachatryan:2015pea}, using the same configuration as in the
SM signal sample.
The 'EWdim6NLO' model~\cite{Alwall:2014hca,Degrande:2012wf} is used for the generation of anomalous couplings.

{\tolerance=2400
For each event, 125 weights are assigned that correspond to a $5{\times} 5{\times} 5$ grid in $c_{WWW}/\Lambda^2 \times c_{W}/\Lambda^2 \times c_{B}/\Lambda^2$.
Equal bins are used in the interval $[-15, 15]\,\text{TeV}^{-2}$ for $c_{WWW}/\Lambda^2$,  $[-50, 50]\,\text{TeV}^{-2}$ for $c_{W}/\Lambda^2$,
and equal bins in the interval $[-500, 500]\,\text{TeV}^{-2}$ for $c_{B}/\Lambda^2$. 
\par}

\subsection{Statistical analysis}\label{atgc:sec5}

The measurement of the coupling constants uses templates in the transverse momentum of the dilepton system
($p_{\mathrm{T} \PZ}$). Because this is well-measured and longitudinally 
Lorentz invariant, this variable is robust against mismodelling and, in principle, ideal for this purpose.
In the electron channel 15 equal bins for $ 0 < p_{\mathrm{T} \PZ} < 900 \GeV $ are used,
and 20 equal bins for $0 < p_{\mathrm{T} \PZ} < 1200 \GeV$ are used in the muon channel, 
where the last bin contains overflow.

\begin{figure*}[htbp]
\centering
\includegraphics[width=\cmsFigWidth]{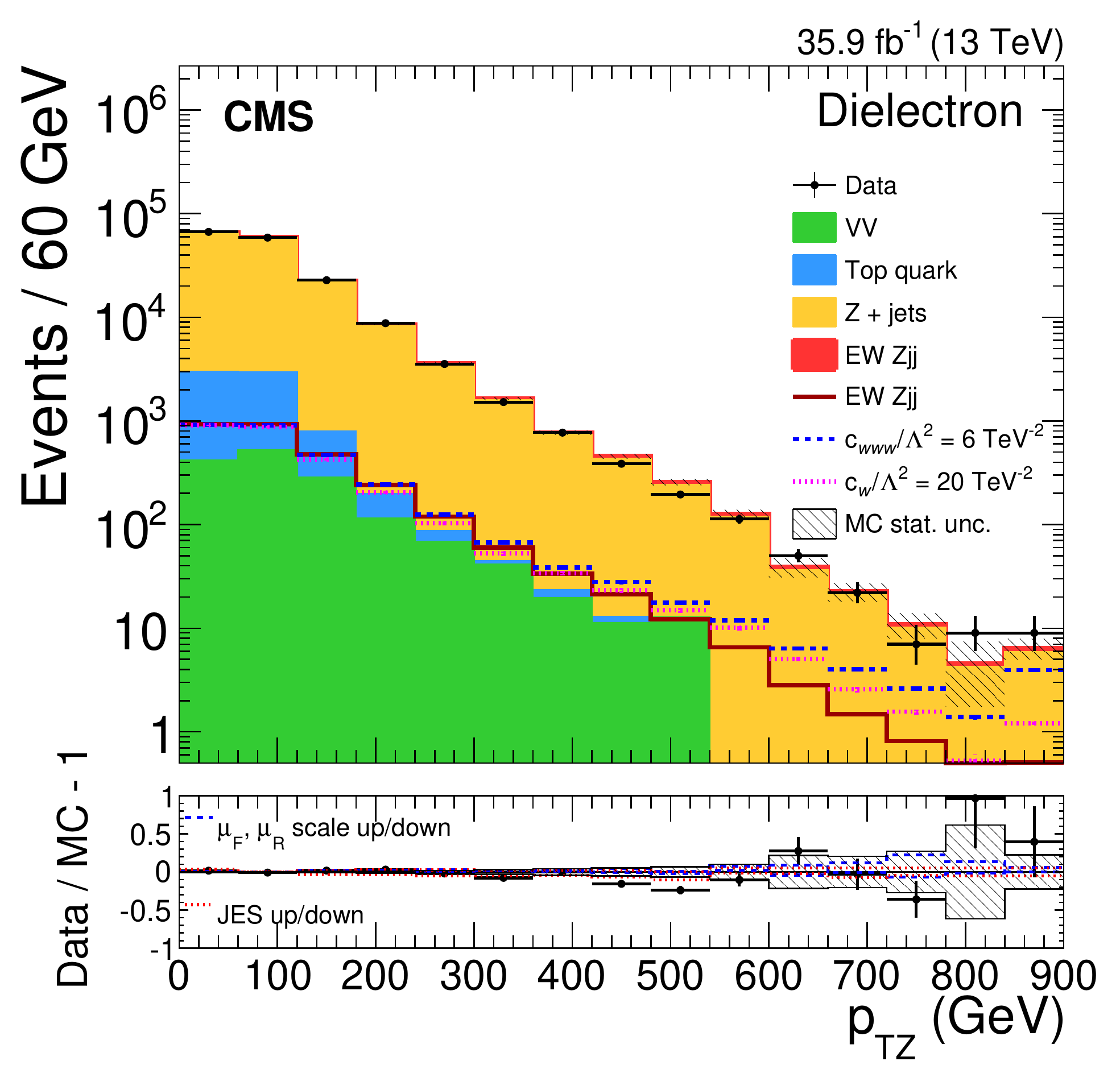}  \hfil
\includegraphics[width=\cmsFigWidth]{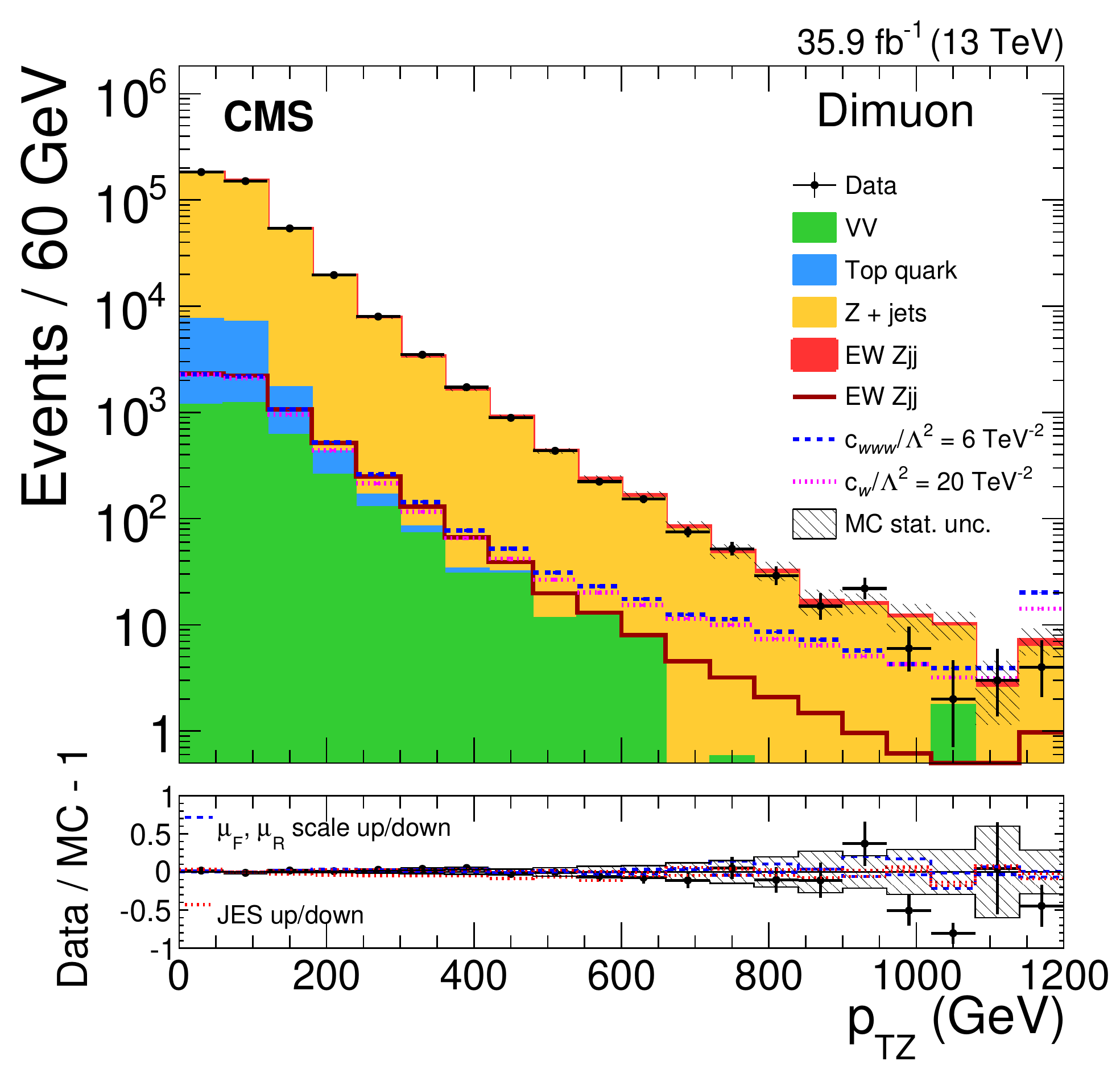}
\caption{\label{atgc:fig5} Distributions of $p_{\mathrm{T} \PZ}$ in data and SM backgrounds, and various
ATGC scenarios in the dielectron (left) and dimuon (right) channels.}
\end{figure*}

In order to construct the $p_{\mathrm{T} \PZ}$ templates, the associated weights calculated for each event are used to construct a parametrized model
of the expected yield in each bin as a function of the values of the dimension-six operators' coupling constants.
For each bin,
the ratios of the expected signal yield with dimension-6 operators to the
one without (leaving only the SM contribution) are fitted at each point of the grid to a quadratic polynomial.
The highest bin is the one with the largest statistical power to detect the presence of higher dimensional operators.
Figure~\ref{atgc:fig5} shows examples of the final templates, with the expected signal overlaid on the background expectation,
for two different hypotheses of dimension-6 operators. The SM distribution is normalized to the expected cross section.

A simultaneous binned fit for the values of the ATGCs is performed in the two lepton channels. 
A profile likelihood method, the Wald Gaussian approximation and Wilks' theorem ~\cite{Cowan:2010js} are used to derive
one-dimensional and two-dimensional
limits at 95\% confidence level (CL) on each of the
three ATGC parameters and each combination of two ATGC parameters, respectively,
while all other parameters are set to their SM values. 
Systematic and theoretical uncertainties are represented by individual nuisance
parameters with log-normal distributions and are profiled in the fit.

\subsection{Results}\label{atgc:sec6}

No significant deviation from the SM expectation is observed. 
Limits on ATGC parameters were previously set by LEP~\cite{Schael:2013ita},
ATLAS~\cite{ATLAS:WV8TeV,Aad:2016ett}, and CMS~\cite{CMS:WV8TeV,Khachatryan:2016poo}.
The LHC semileptonic diboson analyses using 8\TeV data currently set the most stringent limits. 

{\tolerance=600
Limits on the EFT parameters are reported and also translated into the equivalent parameters defined
in an effective Lagrangian (LEP parametrization) in Ref.~\cite{hagiwara1}, without form factors:
$\lambda^{\gamma} = \lambda^{\PZ} = \lambda$, $\Delta{\kappa^{\PZ}} = \Delta{g_1^{\PZ}}-\Delta{\kappa^\gamma} \, \tan^2\theta_{{\PW}}$.
The parameters $\lambda$, $\Delta{\kappa^{Z}}$, and $\Delta{g_1^{\PZ}}$ are considered,
where the $\Delta$ symbols represents deviations from their respective SM values.
\par}

This analysis shows high sensitivity to $c_{WWW}/\Lambda^2$ and $c_{W}/\Lambda^2$ parameters
(equivalently $\lambda^{Z}$ and $\Delta g_{1}^{Z}$).
The sensitivity to $c_{B}/\Lambda^2$ (equivalently $\Delta \kappa^{Z}$)
parameter is very low since the
contribution of this operator to the WWZ vertex is suppressed  by the weak
mixing angle.

\begin{figure*}
\centering
\includegraphics[width=\cmsFigWidth]{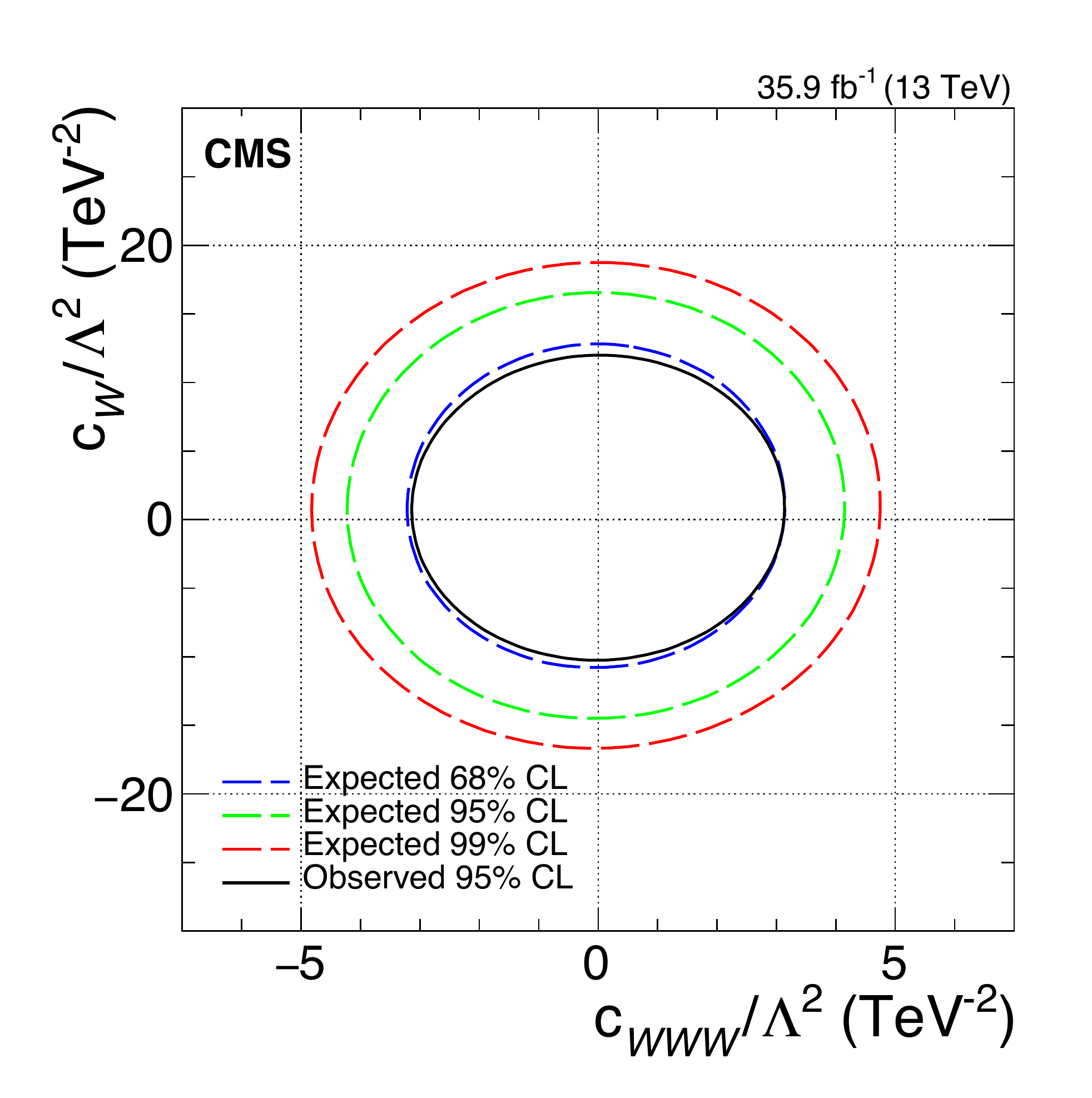} \hfil
\includegraphics[width=\cmsFigWidth]{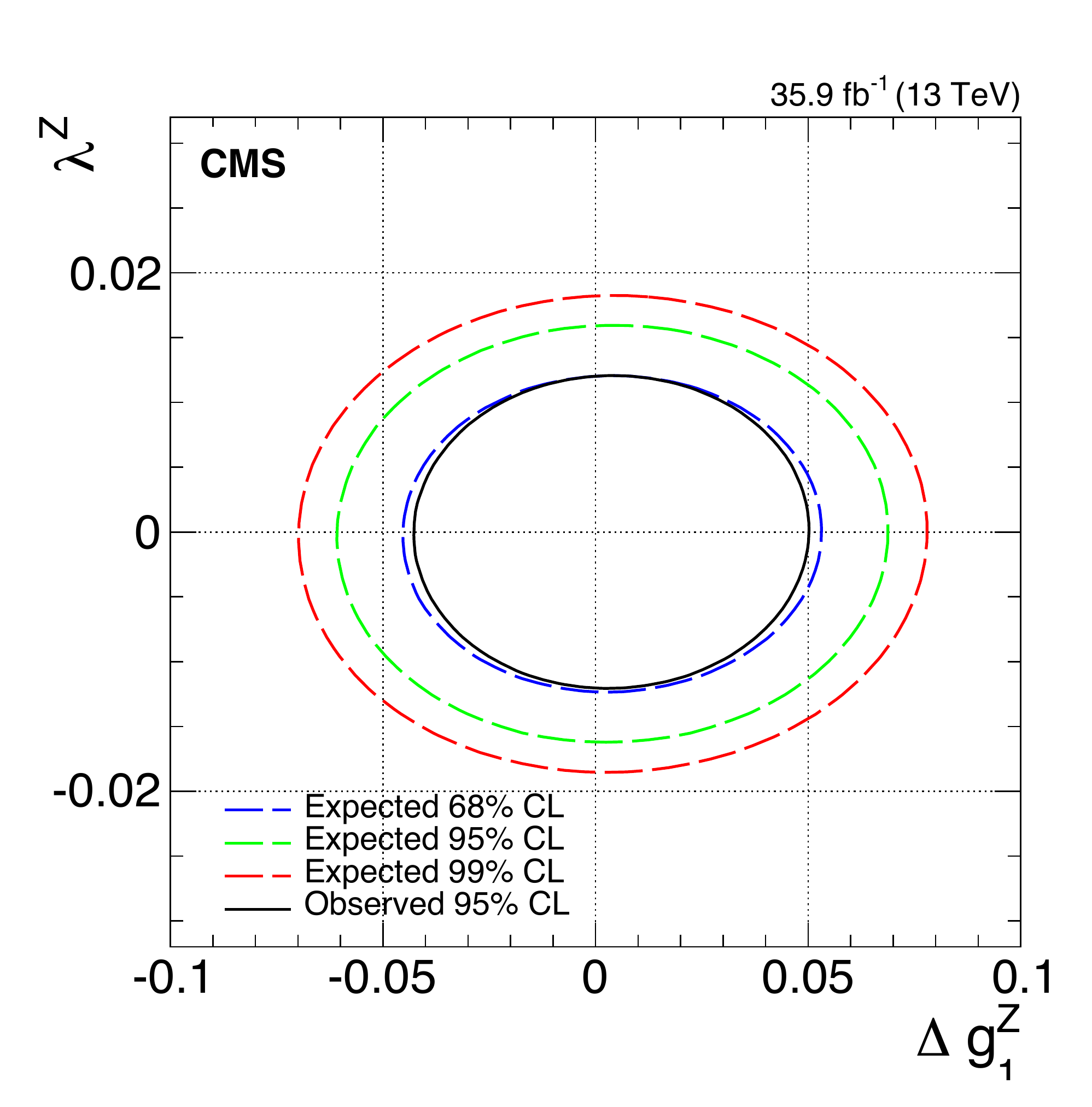}
\caption{\label{atgc:2dlimits_eft}
Two-dimensional observed 95\% CL limits (continuous black line)
and expected 68\%, 95\%, and 99\% CL limits on anomalous coupling parameters.
}
\end{figure*}

Results for 1D limits on $c_{WWW}$ and  $c_W$ ($\lambda$ and $\Delta g_{1}^{\PZ}$) 
can be  found in Table~\ref{atgc:tbl3} (Table~\ref{atgc:limits_atgc}) respectively,
and 2D limits are shown in Fig.~\ref{atgc:2dlimits_eft}.
Results are dominated by the sensitivity in the muon channel
due to the larger acceptance for muons.
An ATGC signal is not included in the interference between EW and DY production.
The effect on the limits is small (${<}3\%$).

\begin{table*}[htbp]
\centering
\topcaption{\label{atgc:tbl3}
One-dimensional limits on the ATGC EFT parameters at 95\% CL.
} 
\renewcommand{\arraystretch}{1.2}
\begin{tabular}{c|c|c}
Coupling constant & Expected 95\% CL interval ($\text{TeV}^{-2}$) & Observed 95\% CL interval ($\text{TeV}^{-2}$)\\ \hline
$c_{WWW}/\Lambda^{2}$ & $[-3.7, 3.6]$ & $[-2.6, 2.6]$\\
$c_{W}/\Lambda^{2}$ & $[-12.6, 14.7]$ & $\x[-8.4, 10.1]$
\end{tabular}
\end{table*}

\begin{table*}[htbp]
\centering
\topcaption{\label{atgc:limits_atgc} 
One-dimensional limits on the ATGC effective Lagrangian (LEP parametrization) parameters at 95\% CL.
} 
\renewcommand{\arraystretch}{1.2}
\begin{tabular}{c|c|c}
Coupling constant & Expected 95\% CL interval & Observed 95\% CL interval \\ \hline
$\lambda^{Z}$ & $[-0.014, 0.014]$ & $[-0.010, 0.010]$\\
$\Delta g_{1}^{Z}$ & $[-0.053, 0.061]$ & $[-0.035, 0.042]$
\end{tabular}
\end{table*}

\section{Study of the hadronic and jet activity in Z + jet events}
\label{sec:hadactivity}

Now that the presence of an SM signal is established, 
the properties of the hadronic
activity in the selected events can be examined.
The production of additional jets in a region with a
larger contribution from \ewkzjj\ processes
is explored in Section~\ref{subsec:highpur}.
Studies of the region in rapidity with expected low hadron
activity (rapidity gap), using track-only observables, are
presented in Section~\ref{subsec:soft}.
Finally a study of hadronic activity vetoes, using both
PF jets and track-only observables, is presented in Section~\ref{subsec:gapveto}.
A significant suppression of the hadronic activity
in signal events is expected because the final-state objects originate
from pure electroweak
interactions, in contrast with the radiative QCD
production of jets in \dyzjj\ events.
The reconstructed distributions are compared directly to the
prediction obtained with a full simulation of the CMS detector.

In the following studies,
event distributions are shown with a selection 
$\mathrm{BDT}>0.92$, which allows a signal-enriched region to be selected
with a similar
fraction of signal and background events.
The $\mathrm{BDT}>0.92$ selection  corresponds approximately to a
selection ${\rm BDT'}>1.946$ on the transformed BDT' discriminants
shown in Fig.~\ref{fig:bdt}.

\subsection{Jet activity studies in a high-purity region}
\label{subsec:highpur}
In this study, aside from the two tagging jets used in the preselection, all PF jets with a
$\pt>15\GeV$ found
within the pseudorapidity gap of the tagging jets,
$\eta^\text{tag jet}_\text{min} < \eta < \eta^\text{tag jet}_\text{max}$,
are used.
The background contribution 
uses the normalizations obtained from the fit discussed in
Section~\ref{sec:results}.

\begin{figure*}[htp]
\centering
\includegraphics[width=\cmsFigWidth]{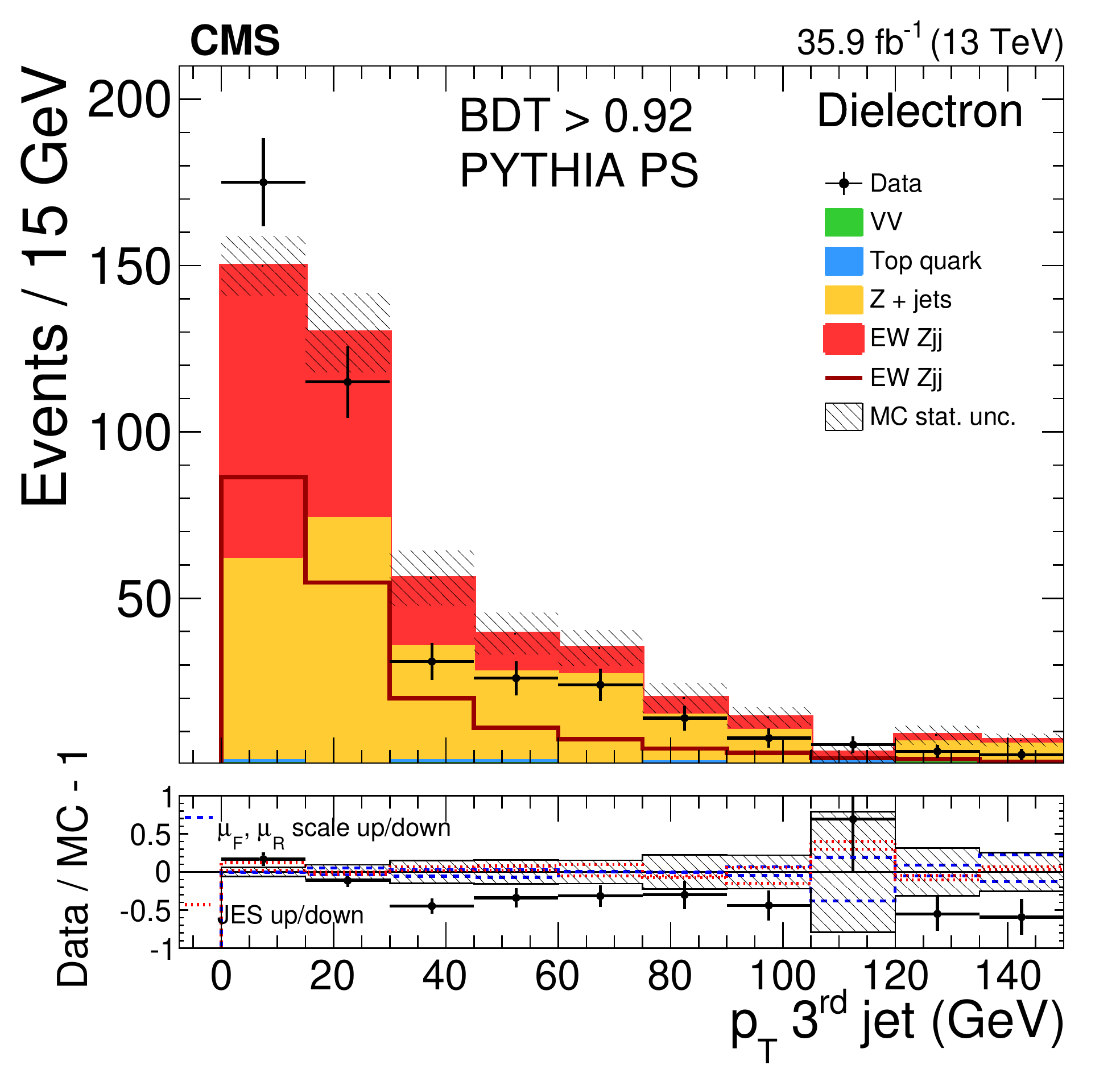} \hfil
\includegraphics[width=\cmsFigWidth]{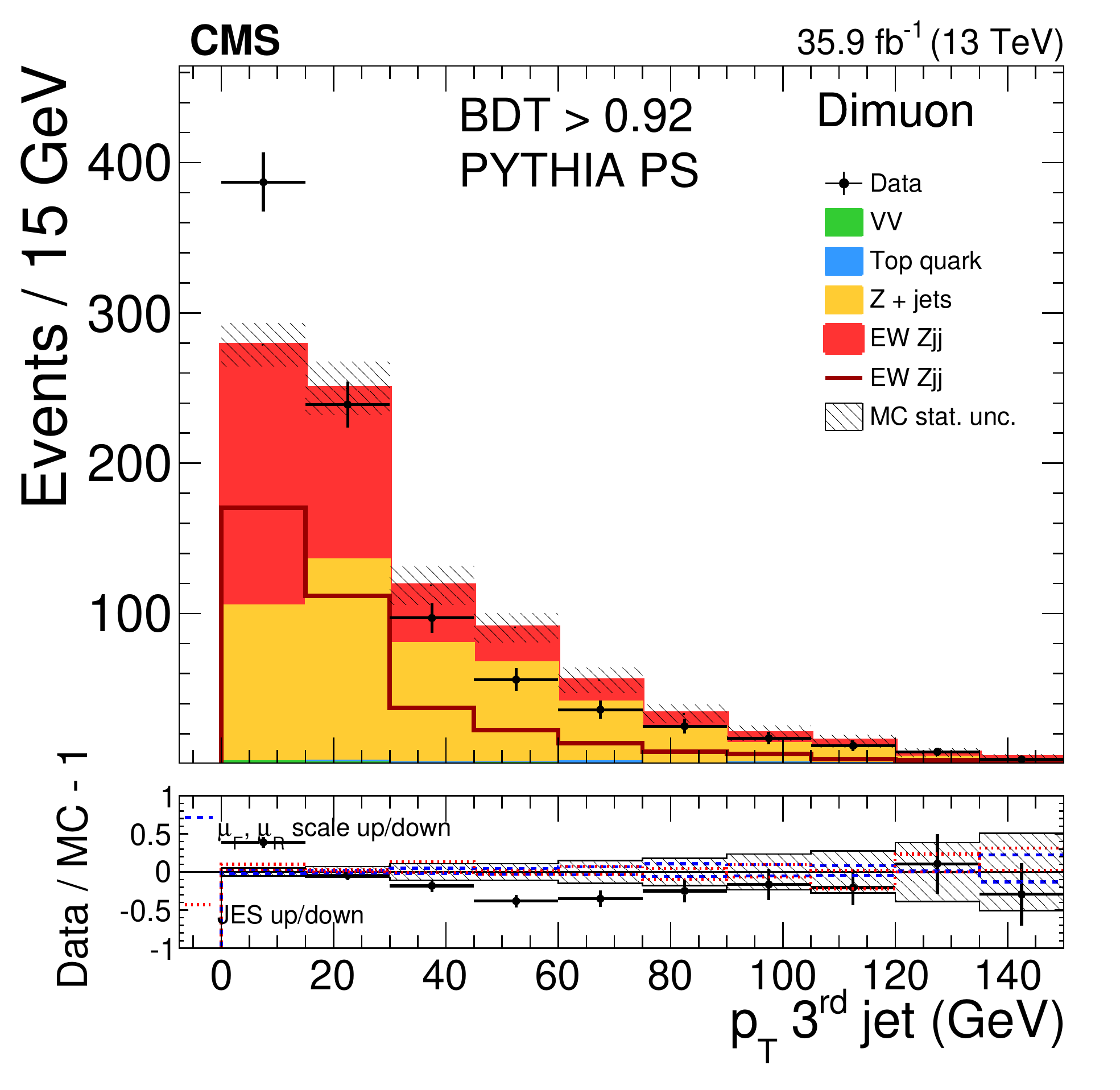} \\
\includegraphics[width=\cmsFigWidth]{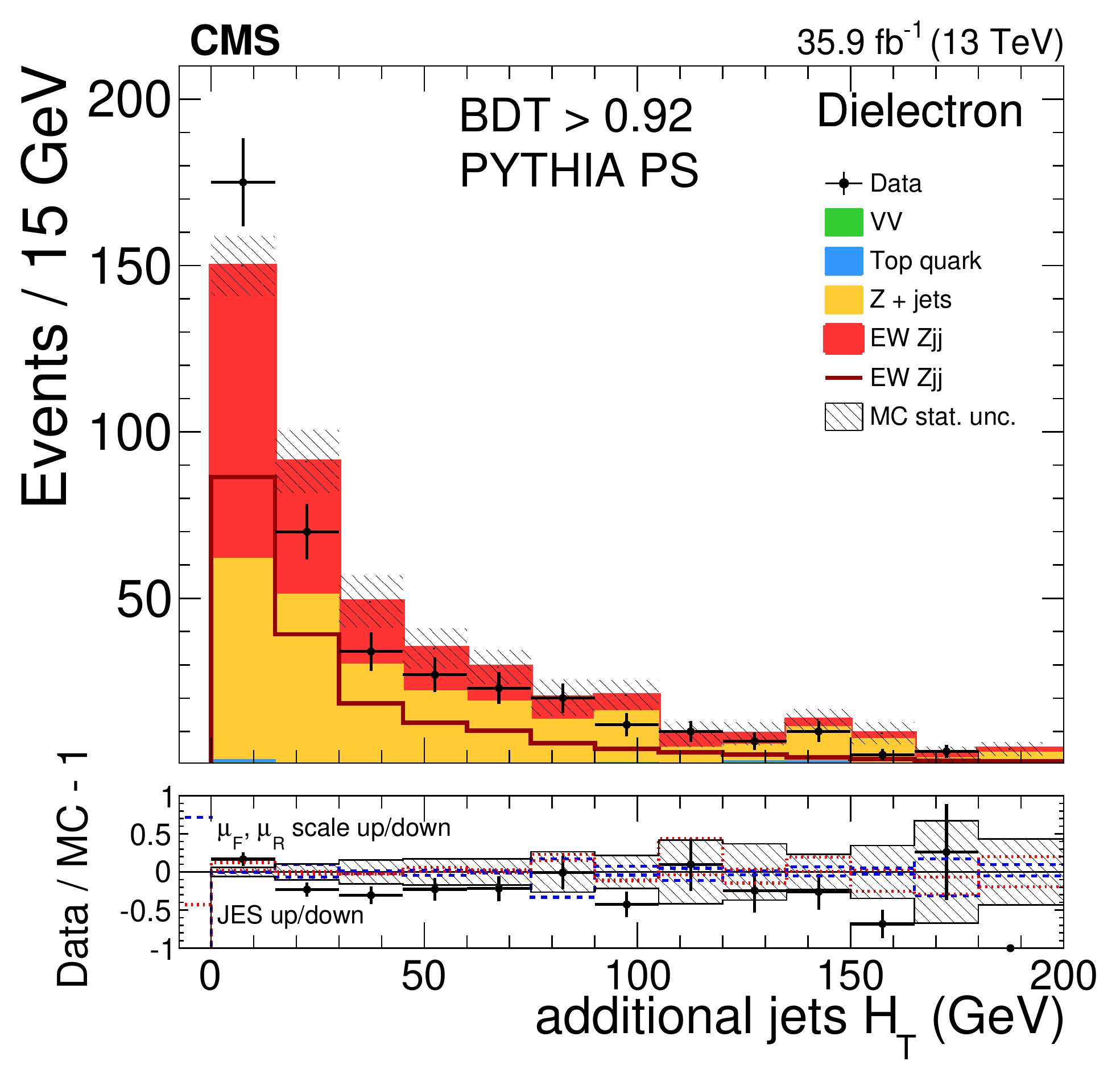} \hfil
\includegraphics[width=\cmsFigWidth]{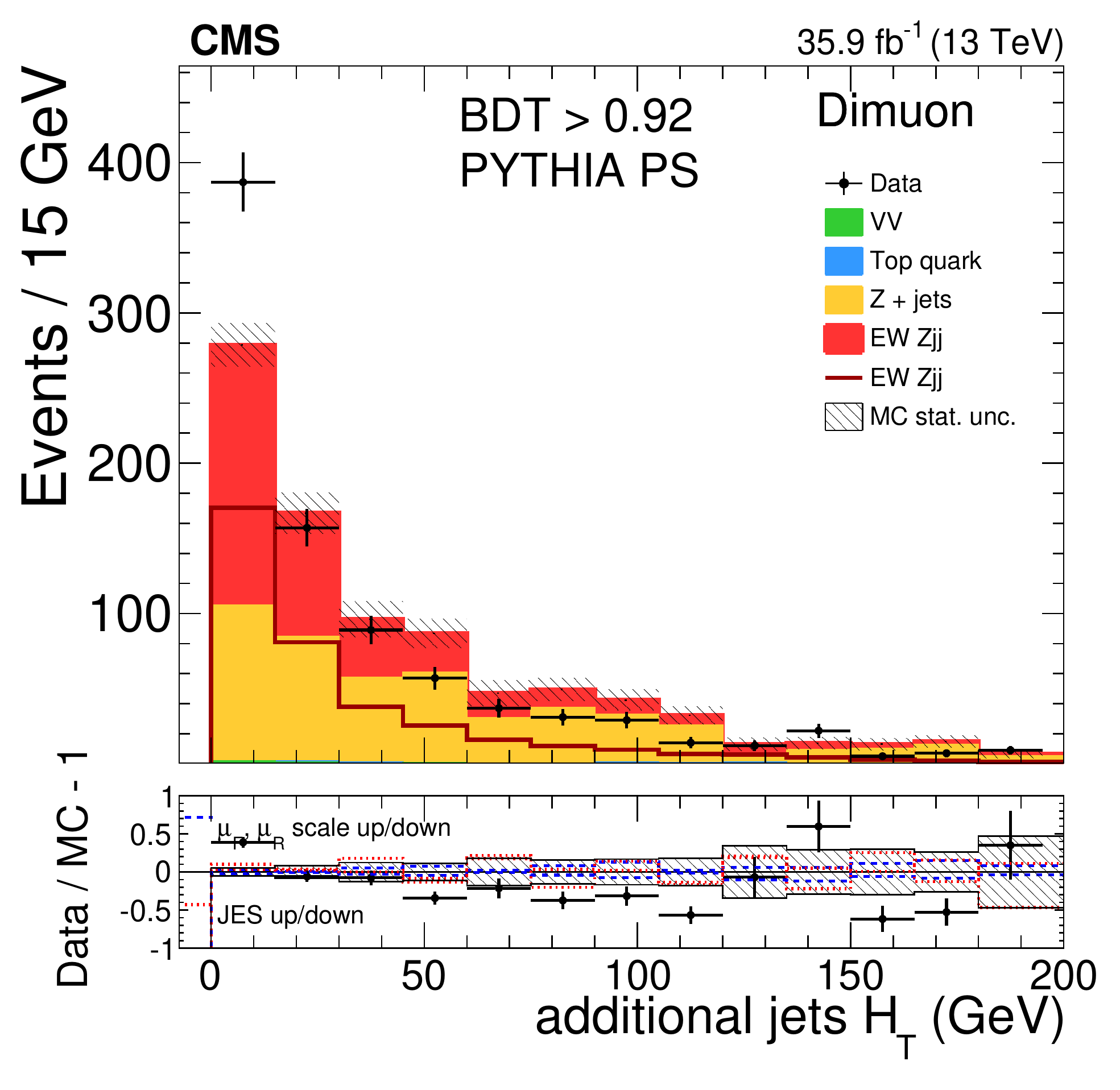}
\caption{
Transverse momentum of the third highest \pt jet  (top row), and
$\HT$ of all additional jets (bottom row)
within the pseudorapidity interval of the two tagging jets
in dielectron (left) and dimuon (right) events with $\mathrm{BDT}>0.92$.
The contributions from the different 
background sources and the signal are shown stacked, with data points superimposed.
The expected signal-only contribution is also shown as an unfilled histogram. 
The lower panels show
the relative difference between the data and expectations,
as well as the uncertainty
envelopes for JES and $\mu_{\rm F,R}$ scale uncertainties.
In all distributions the first bin contains events where no additional jet
with $ \pt>15\GeV$ is present.}
\label{fig:hadingap}
\end{figure*}

The \pt of the \pt-leading additional jet,
as well as the scalar \pt sum ($\HT$) of all additional jets,
are shown in Fig.~\ref{fig:hadingap}.
Data and expectations are generally in reasonable agreement for all
distributions in the signal-enriched regions,
with some deficit of the simulation predictions for the rate of events
with no additional jet activity.
A suppression of the emission of additional jets is observed in
data, when taking into account the background-only predictions.
In the simulation of the signal, the additional jets
are produced by the PS (see Section~\ref{sec:simulation}),
so studying these distributions provides insight on the PS model
in the rapidity-gap region.

\subsection{Study of the charged-hadron activity}
\label{subsec:soft}

For this study, a collection  is formed
of high-purity tracks \cite{CMS-PAS-TRK-10-005} with $\pt > 0.3\GeV$
that are uniquely associated with the main PV in the event.
Tracks associated with the two leptons or with the tagging jets are
excluded from the selection.
The association between the selected tracks and the reconstructed PVs
is carried out by minimizing the longitudinal
impact parameter, which is defined
as the $z$-distance
between the PV and the point of closest approach of
the track helix to the PV, labelled $d_z^\mathrm{PV}$.
The association is required to satisfy $d_z^\mathrm{PV}<2\unit{mm}$ and
$d_z^\mathrm{PV}<3\delta d_z^\mathrm{PV}$, where $\delta d_z^\mathrm{PV}$ is the uncertainty
in $d_z^\mathrm{PV}$.

A collection of ``soft track jets'' is defined
by clustering the selected tracks using the anti-\kt clustering algorithm~\cite{Cacciari:2008gp}
with a distance parameter of $R=0.4$. The use of track jets represents a
clean and well-understood
method~\cite{CMS-PAS-JME-10-006} to reconstruct jets with energy
as low as a few \GeV.
These jets are not affected by pileup because of the association 
of the constituent tracks with the hard-scattering vertex~\cite{CMS-PAS-JME-08-001}.

\begin{figure*}[htp]
\centering
\includegraphics[width=\cmsFigWidth]{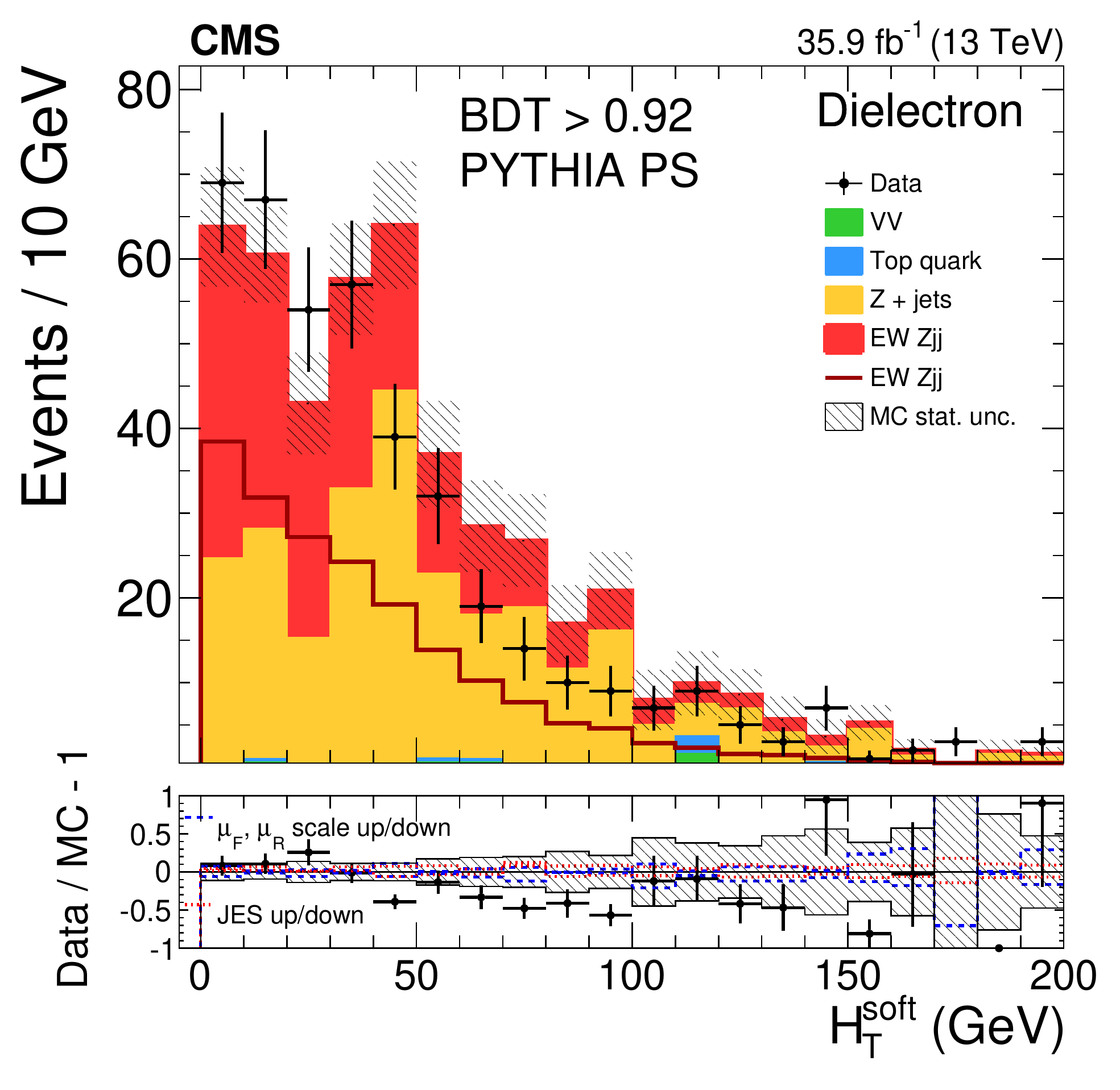} \hfil
\includegraphics[width=\cmsFigWidth]{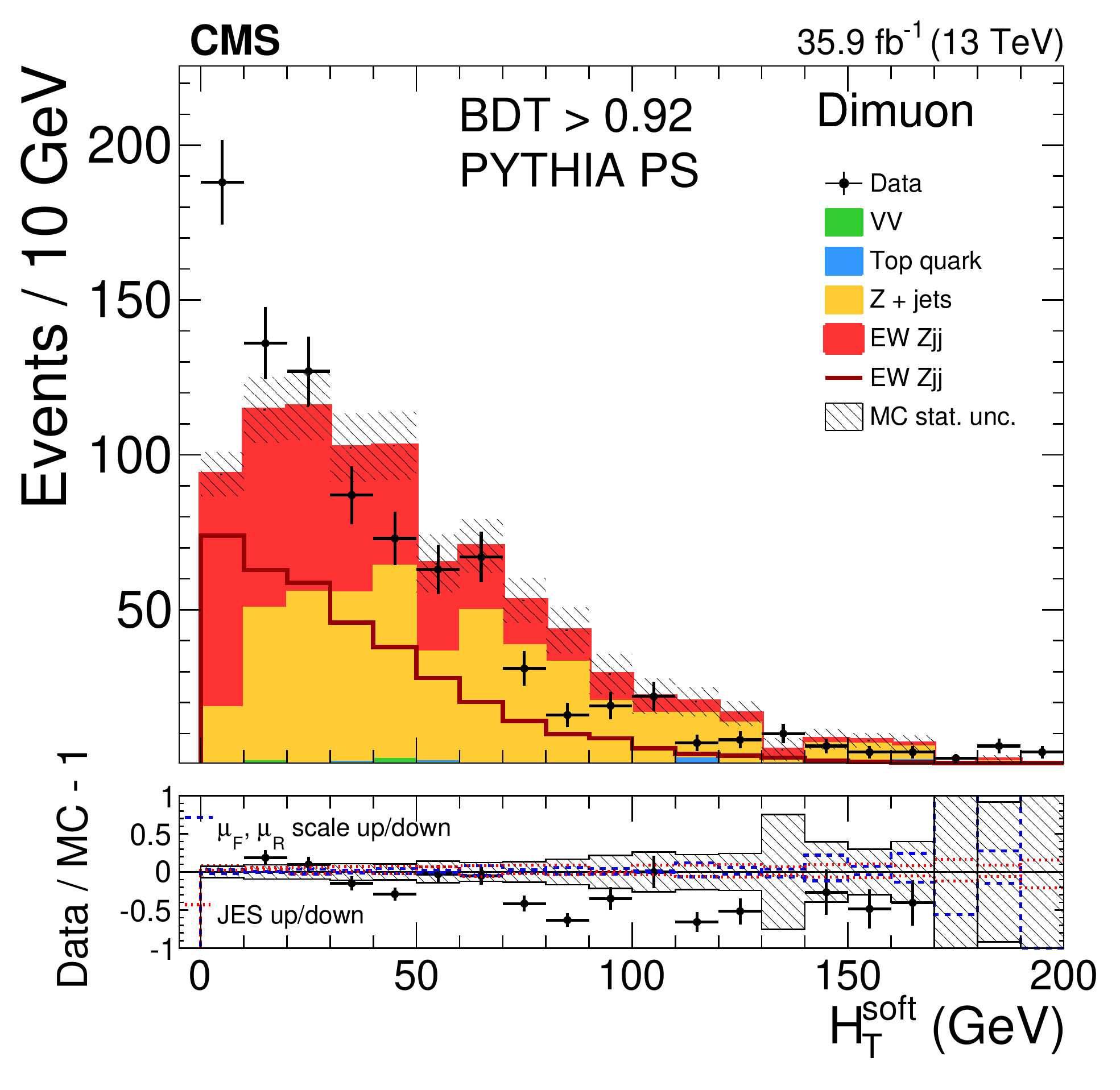} \\
\includegraphics[width=\cmsFigWidth]{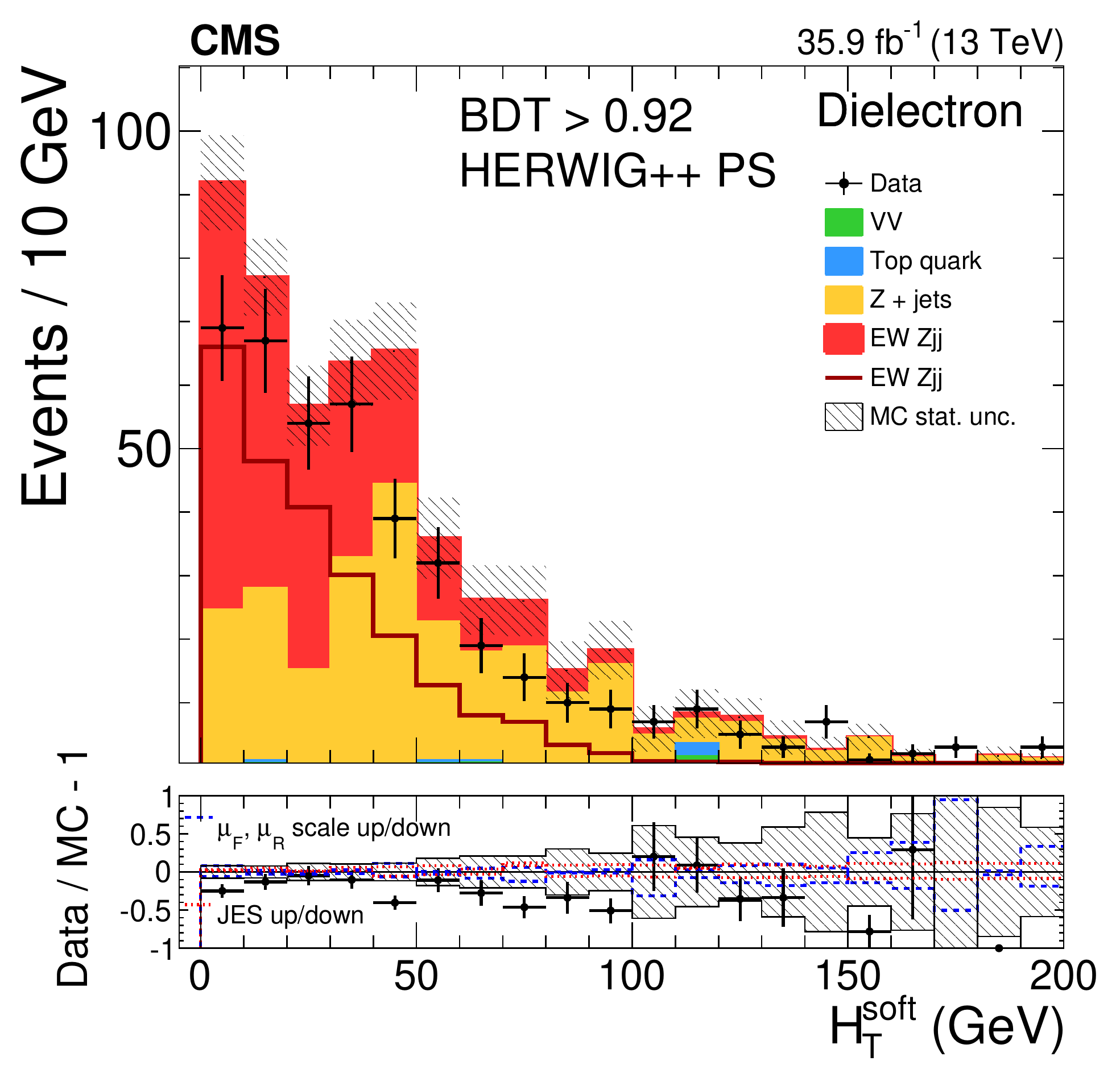} \hfil
\includegraphics[width=\cmsFigWidth]{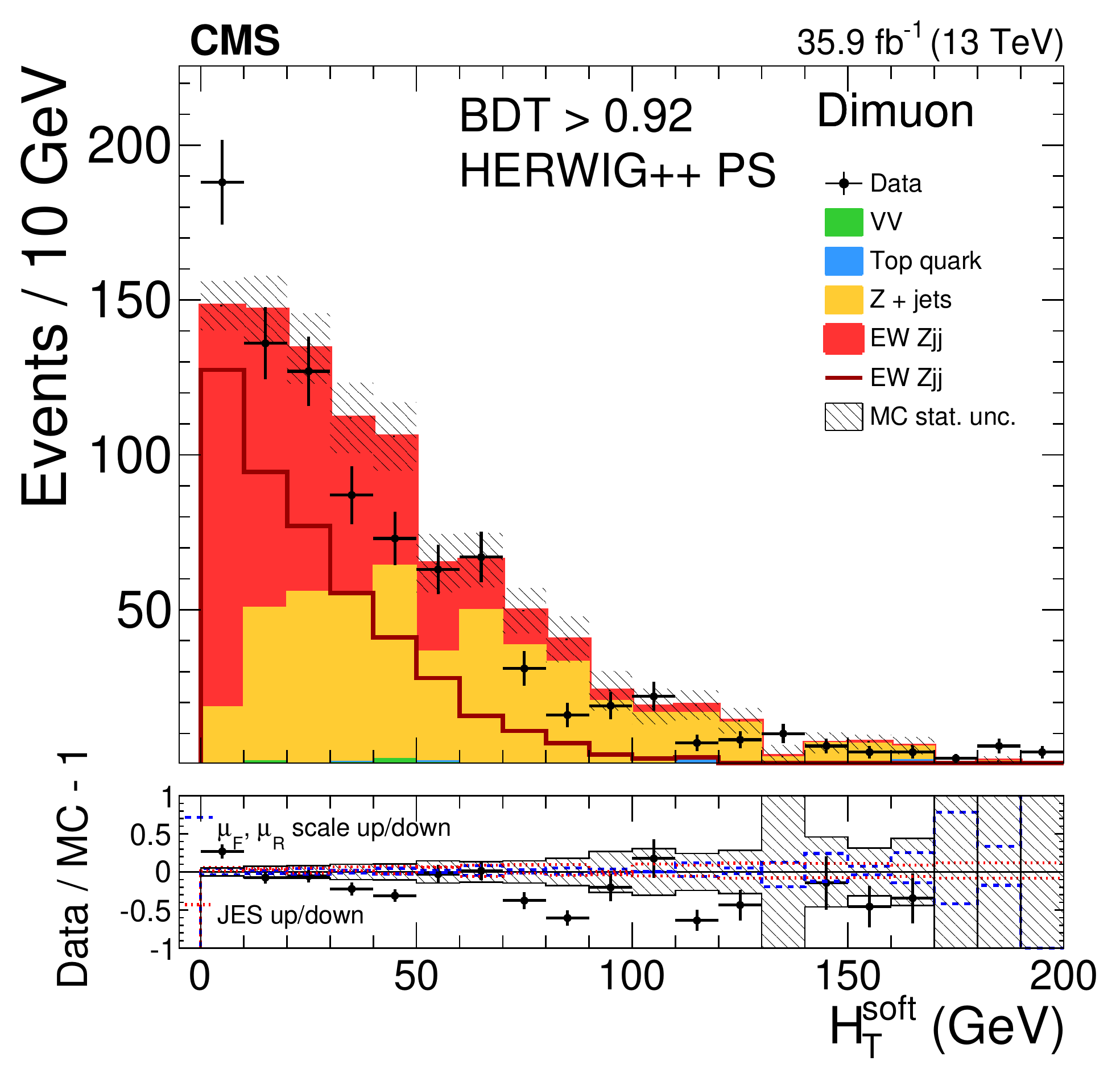}
\caption{
$\HT$ of additional soft track-jets with $\pt>1\GeV$
in dielectron (left) and dimuon (right) events with $\mathrm{BDT}>0.92$.
Data are compared to MC expectations with
the \PYTHIA PS model (top row), or the \HERWIGpp PS model
(bottom row).
The contributions from the different 
background sources and the signal are shown stacked, with data points superimposed.
The expected signal-only contribution is also shown as an unfilled histogram. 
The lower panels show
the relative difference between the data and expectations,
as well as the uncertainty
envelopes for JES and $\mu_{\rm F,R}$ scale uncertainties.}
\label{fig:soft}
\end{figure*}

Track jets of low \pt and within
$\eta^\text{tag jet}_\text{min} < \eta < \eta^\text{tag jet}_\text{max} $ are
considered for the study of the central hadronic activity between the tagging jets.
For each event, the scalar  \pt sum of 
the soft-track jets with $\pt>1\GeV$ is computed, and referred to as
``soft $\HT$''.
Figure~\ref{fig:soft} shows the distribution of the soft \HT
in the signal-enriched region ($\mathrm{BDT}>0.92$), for the dielectron and dimuon
channels, compared to predictions from \PYTHIA and \HERWIGpp PS models.

Overall, a reasonable agreement is observed between data and the simulation.

\subsection{Study of gap activity vetoes}
\label{subsec:gapveto}

The efficiency of a gap activity veto corresponds to the fraction of events with a measured gap activity
below a given threshold. This efficiency can be studied as a function of the applied threshold,
and for different gap activity observables.
The veto thresholds studied here start at 15\GeV for gap activities measured with standard PF jets,
while they go down to 1\GeV for gap activities measured with soft track jets.

\begin{figure*}[htp]
\centering
\includegraphics[width=\cmsFigWidth]{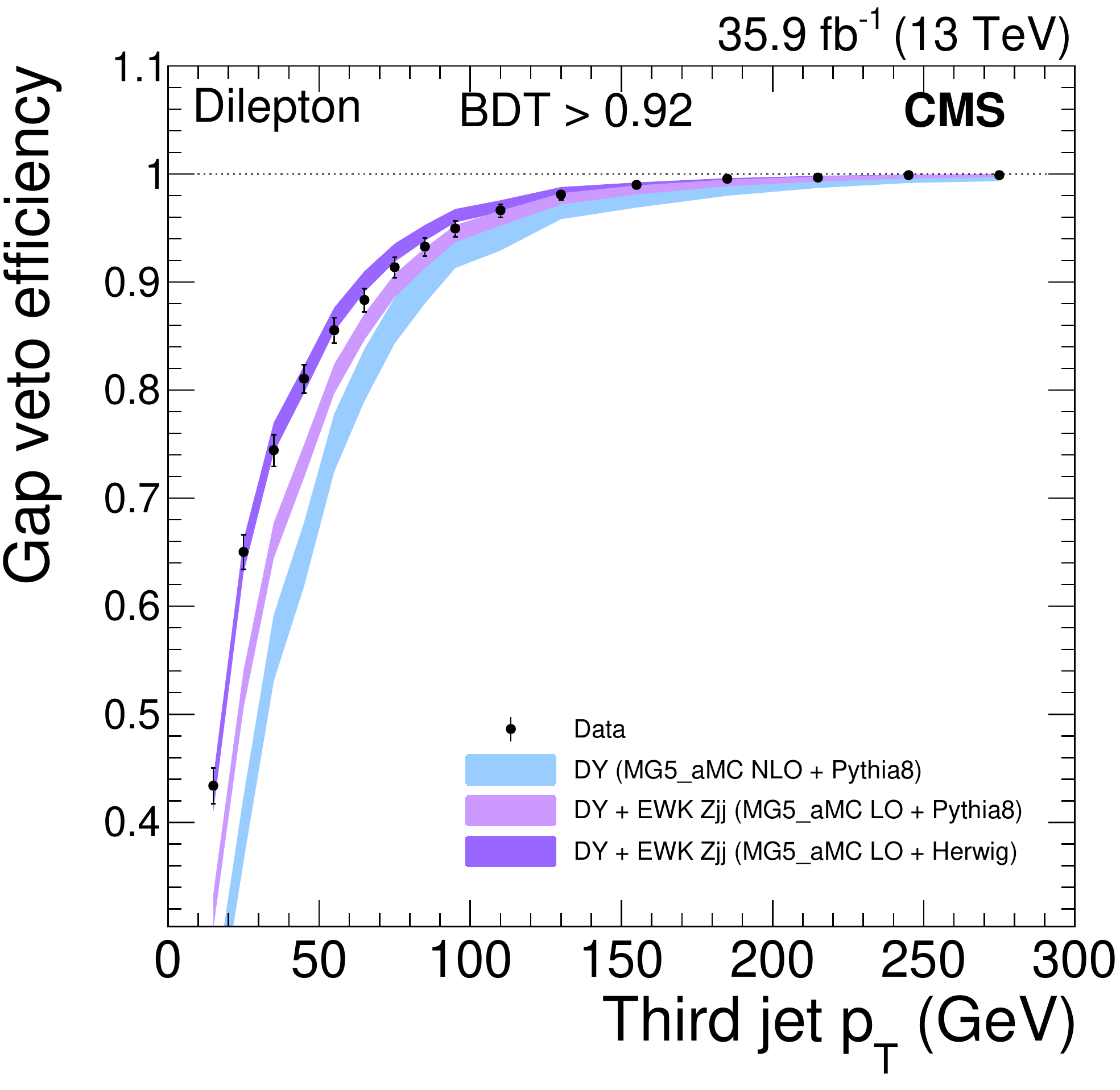} \hfil
\includegraphics[width=\cmsFigWidth]{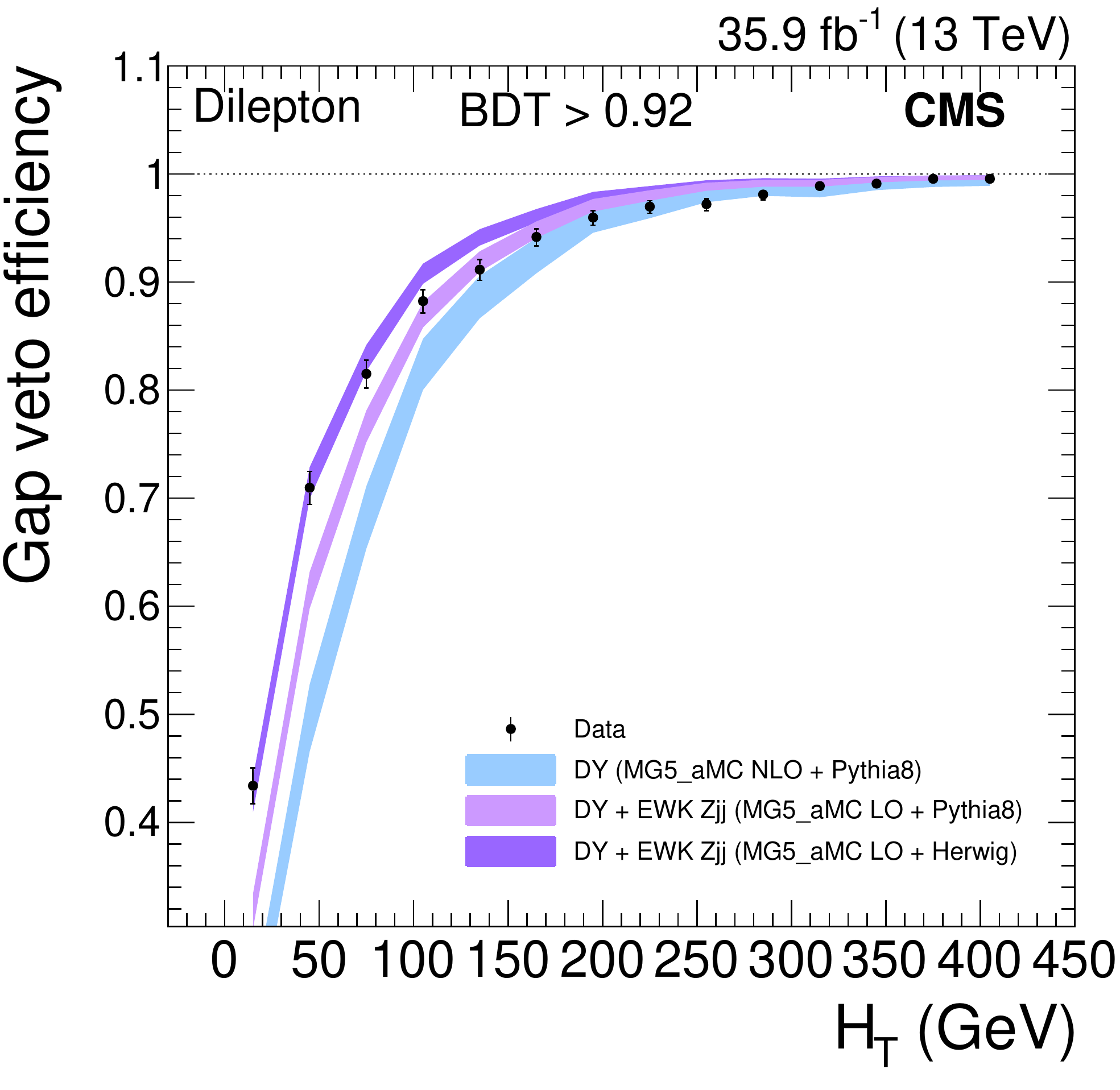}
\caption{
Efficiency of a gap activity veto 
in dielectron and dimuon events with $\mathrm{BDT}>0.92$,
as a function of the additional jet \pt (left),
and of the total \HT of additional jets (right). 
Data points are compared to MC expectations with only DY events,
including signal with the \PYTHIA PS model, or the \HERWIGpp PS model.
The bands represent the MC statistical uncertainty.}
\label{fig:jetveto}
\end{figure*}

Figure~\ref{fig:jetveto} shows the gap activity veto efficiency of combined dielectron and dimuon events
in the signal-enriched region when placing an upper threshold on
the \pt of the additional third jet, or on the total \HT of all additional jets.
The observed efficiency in data is compared to expected efficiencies for
background-only events, and efficiencies for background plus signal events where
the signal is modeled with \PYTHIA or \HERWIGpp.
Data points disfavour the background-only predictions and are in reasonable agreement
with the presence of the signal for both PS predictions. 

\begin{figure*}[htp]
\centering
\includegraphics[width=\cmsFigWidth]{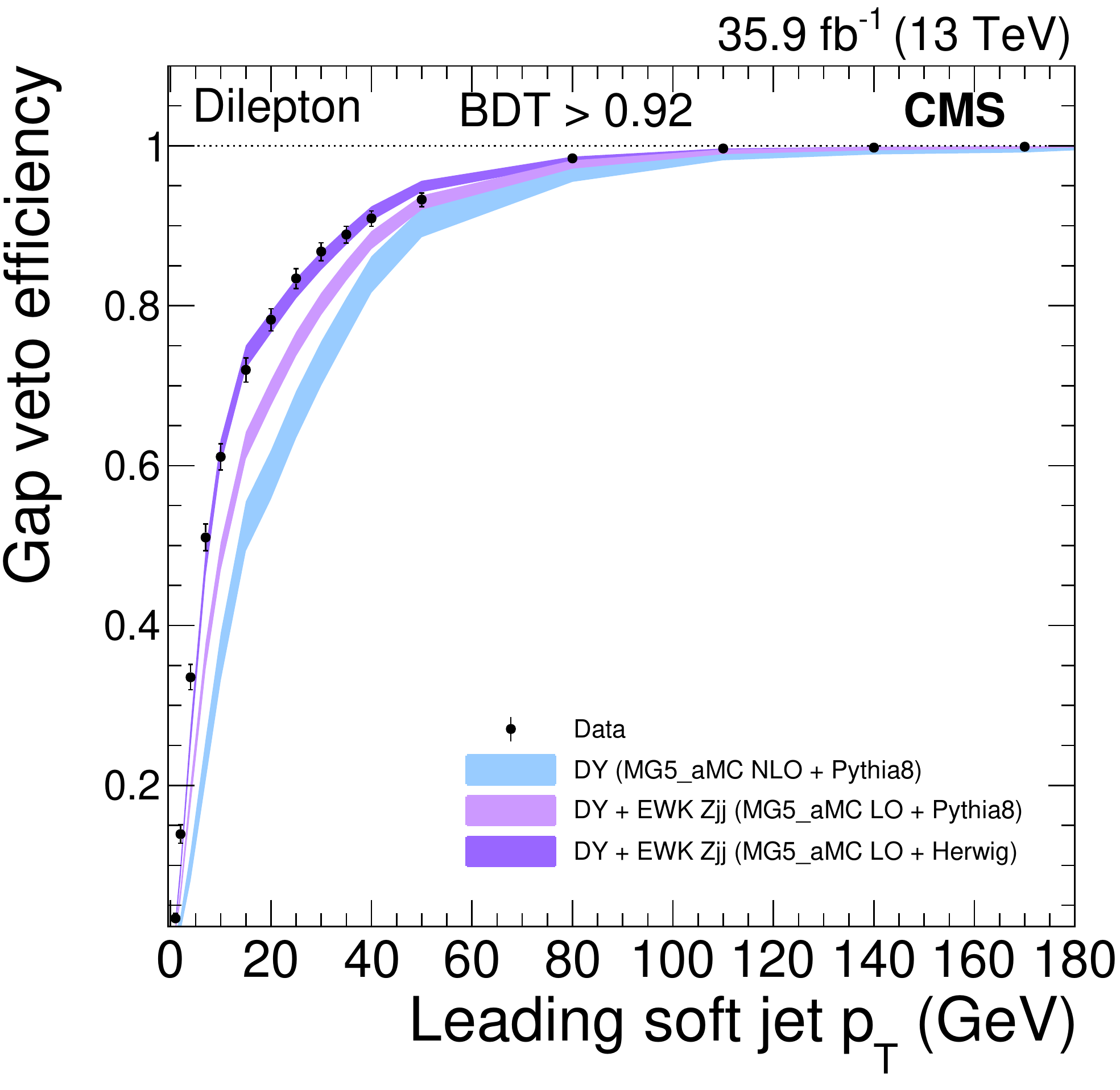} \hfil
\includegraphics[width=\cmsFigWidth]{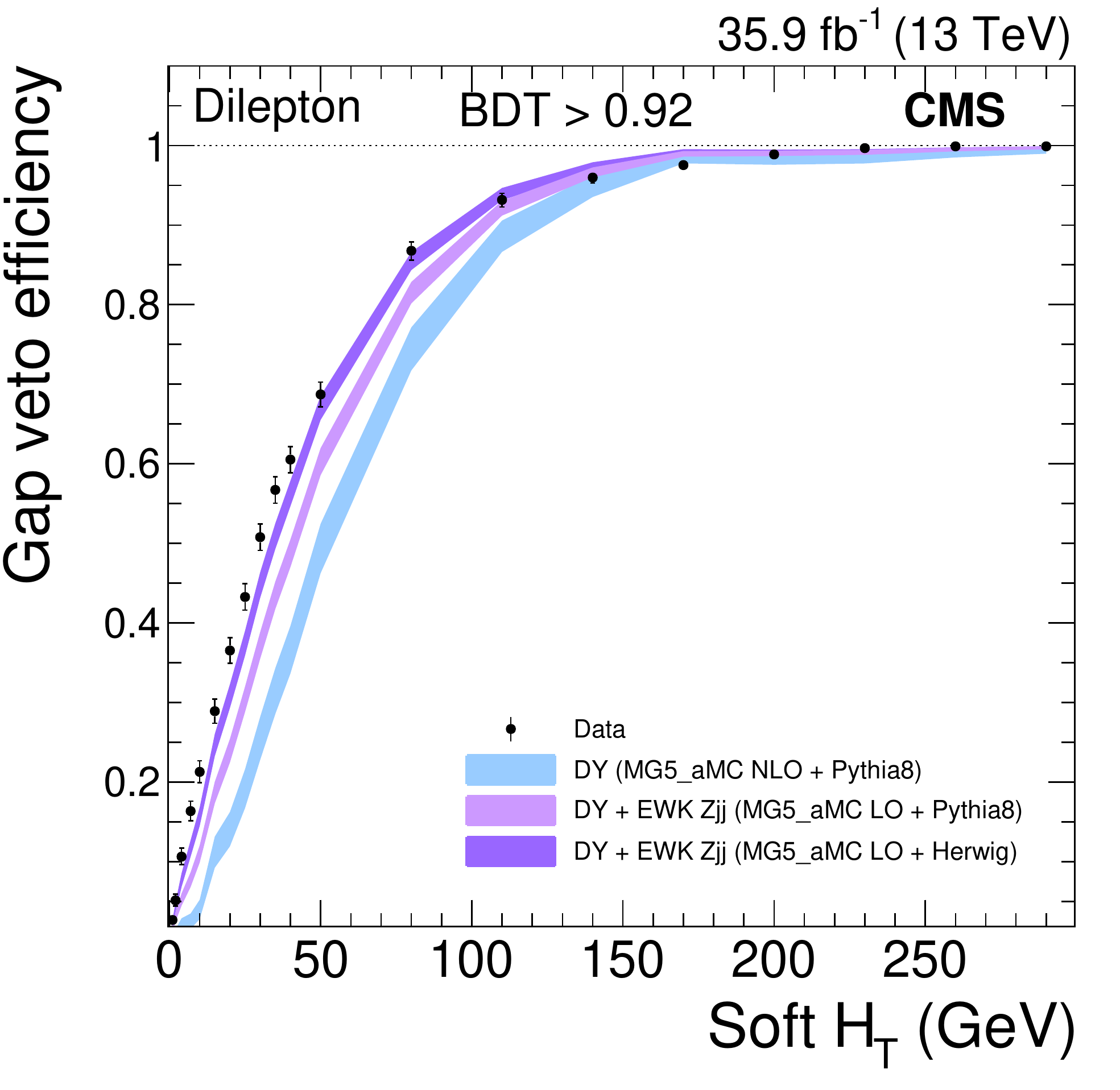}
\caption{
Efficiency of a gap activity veto 
in dielectron and dimuon events with $\mathrm{BDT}>0.92$,
as a function of the leading soft track-jet \pt (left),
and of the total soft \HT (right). 
Data points are compared to MC expectations with only DY events,
including signal with the \PYTHIA PS model, or the \HERWIGpp PS model.
The bands represent the MC statistical uncertainty.}
\label{fig:softveto}
\end{figure*}

Figure~\ref{fig:softveto} shows the gap activity veto efficiency of combined dielectron and dimuon events
in the signal-enriched region when placing an upper threshold on
the \pt of the leading soft jet, or on the total soft \HT.
The data points disfavour the background-only predictions and are in reasonable agreement
with the presence of the signal with both PS predictions.
Comparisons between the signal gap activity predictions obtained
with \PYTHIA PS model and the \HERWIGpp PS model have been previously studied~\cite{Schissler:2013nga}, and are
consistent with the predictions found here.
Among the two considered signal models, the data seem to prefer
the signal model with \HERWIGpp PS at low gap activity values,
whereas the \PYTHIA (v8.212) PS predictions seem to be preferred 
by the data in the case of larger gap activities. 

\section{Summary}
\label{sec:summary}

The cross section for the electroweak (EW) production of a Z boson
in association with two jets
in the \lljj\ final state is measured in  proton-proton collisions at
$\sqrt{s}=13\TeV$
in the kinematic region defined by
$m_{\ell\ell} >50\GeV$, $m_{\mathrm{jj}} >120\GeV$,
and transverse momenta $p_\mathrm{T j} > 25\GeV$.
The result
\begin{equation*}
\sigma({\mathrm{EW}~\ell\ell\mathrm{jj}})=534\pm 20\stat \pm 57\syst\unit{fb},
\end{equation*}
agrees with the standard model prediction.

The increased cross section and integrated luminosity recorded at 13\TeV,
as well as the more precise NLO modelling of background processes,
have led to 
a more precise measurement of the \ewkzjj\ process, relative
to earlier CMS and ATLAS results,
where the relative precision was approximately
20\%~\cite{Chatrchyan:2013jya,Khachatryan:2014dea,Aad:2014dta,Aaboud:2017emo}.

No evidence for anomalous trilinear gauge couplings is found.
The following one-dimensional limits at 95\% CL are obtained: $-2.6 <  c_{WWW}/\Lambda^2  < 2.6\TeV^{-2}$ and 
$-8.4 <  c_{W}/\Lambda^2  < 10.1\TeV^{-2}$.
These results provide the most stringent constraints on $c_{WWW}$ to date.

In events from a signal-enriched region, the additional hadron activity
is also studied,
as well as the efficiencies for a gap-activity veto,
and generally good agreement is found between data and quantum chromodynamics predictions
with either the \PYTHIA or \HERWIGpp
parton shower and hadronization 
model.

\begin{acknowledgments}

\hyphenation{Bundes-ministerium Forschungs-gemeinschaft Forschungs-zentren Rachada-pisek} We congratulate our colleagues in the CERN accelerator departments for the excellent performance of the LHC and thank the technical and administrative staffs at CERN and at other CMS institutes for their contributions to the success of the CMS effort. In addition, we gratefully acknowledge the computing centers and personnel of the Worldwide LHC Computing Grid for delivering so effectively the computing infrastructure essential to our analyses. Finally, we acknowledge the enduring support for the construction and operation of the LHC and the CMS detector provided by the following funding agencies: the Austrian Federal Ministry of Science, Research and Economy and the Austrian Science Fund; the Belgian Fonds de la Recherche Scientifique, and Fonds voor Wetenschappelijk Onderzoek; the Brazilian Funding Agencies (CNPq, CAPES, FAPERJ, and FAPESP); the Bulgarian Ministry of Education and Science; CERN; the Chinese Academy of Sciences, Ministry of Science and Technology, and National Natural Science Foundation of China; the Colombian Funding Agency (COLCIENCIAS); the Croatian Ministry of Science, Education and Sport, and the Croatian Science Foundation; the Research Promotion Foundation, Cyprus; the Secretariat for Higher Education, Science, Technology and Innovation, Ecuador; the Ministry of Education and Research, Estonian Research Council via IUT23-4 and IUT23-6 and European Regional Development Fund, Estonia; the Academy of Finland, Finnish Ministry of Education and Culture, and Helsinki Institute of Physics; the Institut National de Physique Nucl\'eaire et de Physique des Particules~/~CNRS, and Commissariat \`a l'\'Energie Atomique et aux \'Energies Alternatives~/~CEA, France; the Bundesministerium f\"ur Bildung und Forschung, Deutsche Forschungsgemeinschaft, and Helmholtz-Gemeinschaft Deutscher Forschungszentren, Germany; the General Secretariat for Research and Technology, Greece; the National Scientific Research Foundation, and National Innovation Office, Hungary; the Department of Atomic Energy and the Department of Science and Technology, India; the Institute for Studies in Theoretical Physics and Mathematics, Iran; the Science Foundation, Ireland; the Istituto Nazionale di Fisica Nucleare, Italy; the Ministry of Science, ICT and Future Planning, and National Research Foundation (NRF), Republic of Korea; the Lithuanian Academy of Sciences; the Ministry of Education, and University of Malaya (Malaysia); the Mexican Funding Agencies (BUAP, CINVESTAV, CONACYT, LNS, SEP, and UASLP-FAI); the Ministry of Business, Innovation and Employment, New Zealand; the Pakistan Atomic Energy Commission; the Ministry of Science and Higher Education and the National Science Centre, Poland; the Funda\c{c}\~ao para a Ci\^encia e a Tecnologia, Portugal; JINR, Dubna; the Ministry of Education and Science of the Russian Federation, the Federal Agency of Atomic Energy of the Russian Federation, Russian Academy of Sciences, the Russian Foundation for Basic Research and the Russian Competitiveness Program of NRNU ``MEPhI"; the Ministry of Education, Science and Technological Development of Serbia; the Secretar\'{\i}a de Estado de Investigaci\'on, Desarrollo e Innovaci\'on, Programa Consolider-Ingenio 2010, Plan de Ciencia, Tecnolog\'{i}a e Innovaci\'on 2013-2017 del Principado de Asturias and Fondo Europeo de Desarrollo Regional, Spain; the Swiss Funding Agencies (ETH Board, ETH Zurich, PSI, SNF, UniZH, Canton Zurich, and SER); the Ministry of Science and Technology, Taipei; the Thailand Center of Excellence in Physics, the Institute for the Promotion of Teaching Science and Technology of Thailand, Special Task Force for Activating Research and the National Science and Technology Development Agency of Thailand; the Scientific and Technical Research Council of Turkey, and Turkish Atomic Energy Authority; the National Academy of Sciences of Ukraine, and State Fund for Fundamental Researches, Ukraine; the Science and Technology Facilities Council, UK; the US Department of Energy, and the US National Science Foundation.

Individuals have received support from the Marie-Curie program and the European Research Council and Horizon 2020 Grant, contract No. 675440 (European Union); the Leventis Foundation; the A. P. Sloan Foundation; the Alexander von Humboldt Foundation; the Belgian Federal Science Policy Office; the Fonds pour la Formation \`a la Recherche dans l'Industrie et dans l'Agriculture (FRIA-Belgium); the Agentschap voor Innovatie door Wetenschap en Technologie (IWT-Belgium); the Ministry of Education, Youth and Sports (MEYS) of the Czech Republic; the Council of Scientific and Industrial Research, India; the HOMING PLUS program of the Foundation for Polish Science, cofinanced from European Union, Regional Development Fund, the Mobility Plus program of the Ministry of Science and Higher Education, the National Science Center (Poland), contracts Harmonia 2014/14/M/ST2/00428, Opus 2014/13/B/ST2/02543, 2014/15/B/ST2/03998, and 2015/19/B/ST2/02861, Sonata-bis 2012/07/E/ST2/01406; the National Priorities Research Program by Qatar National Research Fund; the Programa Severo Ochoa del Principado de Asturias; the Thalis and Aristeia programs cofinanced by EU-ESF and the Greek NSRF; the Rachadapisek Sompot Fund for Postdoctoral Fellowship, Chulalongkorn University and the Chulalongkorn Academic into Its 2nd Century Project Advancement Project (Thailand); the Welch Foundation, contract C-1845; and the Weston Havens Foundation (USA).

\end{acknowledgments}

\bibliography{auto_generated}

\cleardoublepage \appendix\section{The CMS Collaboration \label{app:collab}}\begin{sloppypar}\hyphenpenalty=5000\widowpenalty=500\clubpenalty=5000\vskip\cmsinstskip
\textbf{Yerevan Physics Institute,  Yerevan,  Armenia}\\*[0pt]
A.M.~Sirunyan,  A.~Tumasyan
\vskip\cmsinstskip
\textbf{Institut f\"{u}r Hochenergiephysik,  Wien,  Austria}\\*[0pt]
W.~Adam,  F.~Ambrogi,  E.~Asilar,  T.~Bergauer,  J.~Brandstetter,  E.~Brondolin,  M.~Dragicevic,  J.~Er\"{o},  A.~Escalante Del Valle,  M.~Flechl,  M.~Friedl,  R.~Fr\"{u}hwirth\cmsAuthorMark{1},  V.M.~Ghete,  J.~Grossmann,  J.~Hrubec,  M.~Jeitler\cmsAuthorMark{1},  A.~K\"{o}nig,  N.~Krammer,  I.~Kr\"{a}tschmer,  D.~Liko,  T.~Madlener,  I.~Mikulec,  E.~Pree,  N.~Rad,  H.~Rohringer,  J.~Schieck\cmsAuthorMark{1},  R.~Sch\"{o}fbeck,  M.~Spanring,  D.~Spitzbart,  A.~Taurok,  W.~Waltenberger,  J.~Wittmann,  C.-E.~Wulz\cmsAuthorMark{1},  M.~Zarucki
% ...
\vskip\cmsinstskip
% ...
1:~Also at Vienna University of Technology,  Vienna,  Austria\\
% ...
\end{sloppypar}
\end{document}